\documentclass[utf8]{FrontiersinHarvard} 

\usepackage{url,hyperref,lineno,microtype,subcaption}
\usepackage{graphicx}
\usepackage{revsymb}	
\usepackage{amsmath}	
\usepackage{hyperref} 
\usepackage[onehalfspacing]{setspace}
\usepackage{bm}
\nolinenumbers

\def\keyFont{\fontsize{8}{11}\helveticabold }
\def\firstAuthorLast{Buldgen {et~al.}} 
\def\Authors{Gaël Buldgen\,$^{1,*}$, Jérôme Bétrisey\,$^{1}$, Ian W. Roxburgh\,$^{2,3}$, Sergei V. Vorontsov\,$^{2,4}$ and Daniel R. Reese\,$^{5}$}
\def\Address{$^{1}$ D\'epartement d'Astronomie, Universit\'e de Gen\`eve, Chemin Pegasi 51, 1290, Sauverny, Switzerland\\
$^{2}$ Astronomy Unit, Queen Mary University of London, Mile End Road, London, E1 4NS, UK\\
$^{3}$ School of Physics and Astronomy, University of Birmingham, Edgbaston, Birmingham, B15 2TT, UK\\
$^{4}$ Institute of Physics of the Earth, B. Gruzinskaya 10, Moscow 123995, Russia\\
$^{5}$ LESIA, Observatoire de Paris, Universit\'e PSL, CNRS, Sorbonne Universit\'e, Universit\'e Paris Cit\'e, 5 place Jules Janssen, 92195, Meudon, France}
\def\corrAuthor{Gaël Buldgen}
\def\corrEmail{Gael.Buldgen@unige.ch}

\begin{document}
\onecolumn
\firstpage{1}

\title[Inversions of stellar structure]{Inversions of stellar structure from asteroseismic data} 

\author[\firstAuthorLast ]{\Authors} 
\address{} 
\correspondance{} 

\extraAuth{}

\maketitle

\begin{abstract}

The advent of space-based photometry missions in the early 21st century enabled the application to asteroseismic data of advanced inference techniques until then restricted to the field of helioseismology. The high quality of the observations, the discovery of mixed modes in evolved solar-like oscillators and the need for an improvement in the determination of stellar fundamental parameters such as mass, radius and age led to the development of sophisticated modelling tools, amongst which seismic inversions play a key role. In this review, we will discuss the existing inversion techniques for the internal structure of distant stars adapted from helio- to asteroseismology. We will present results obtained for various Kepler targets, their coupling to other existing modelling techniques as well as the limitations of seismic analyses and the perspectives for future developments of these approaches in the context of the current TESS and the future PLATO mission, as well as the exploitation of the mixed modes observed in post-main sequence solar-like oscillators, for which variational formulations might not provide sufficient accuracy. 
\tiny
 \keyFont{ \section{Keywords:} Asteroseismology, Stellar Physics, Inversion techniques, Solar-like Stars, Stellar Evolution} 
\end{abstract}

\section{Introduction}

In the last two decades, asteroseismology has established itself as the golden path to study the internal structure of distant stars. Amongst these, solar-like oscillators have held a special place due to the number of them detected at various masses, chemical composition and evolutionary stages. This revolution was made possible by the recent space-based photometry missions CoRoT \citep{Baglin}, \textit{Kepler} \citep{Borucki} and TESS \citep{Ricker2015}. In the near future, the PLATO mission \citep{Rauer2014} will further extend the dataset, and proposals for future missions have also been laid out, testifying to the scientific success of these missions \citep[e.g.][]{Miglio2021}. 

In addition to providing numerous targets to work on, space-based photometry missions also led to a drastic change in data quality compared to previous ground-based observations. While ground-based telescope networks \citep[see][for the SONG network]{Grundahl2006} are still very important and can provide high quality data for some close targets \citep[e.g.][]{Bouchy2001,Martic2004,Bedding2004,Kjeldsen2005A,Grundahl2006}, they are no match for nanohertz precision on the observed frequencies of some of the best \textit{Kepler} targets and the large number of stars that can be observed simultaneously by photometric surveys. 

This drastic change motivated the use of advanced seismic analyses techniques that were before restricted to the field of helioseismology, where the proximity of the Sun allows the detection of thousands of oscillation modes. Seismic inferences became routinely used and the determination of fundamental parameters of stars for the purposes of fields such as exoplanetology and Galactic archaeology drove the development of dedicated numerical tools. Large grids of stellar models were also computed for such purposes and coupled to automated seismic modelling pipelines \citep[e.g.][]{Mathur2012,Gruberbauer2013,Metcalfe2015,Bellinger2016,Rendle2019,Bazot2020,Silva2022} .  

Inference problems are widespread in physics, from the calibration of instrumental responses to the radiative transfer problem in the Earth's atmosphere. They correspond to a specific class of mathematical problems called ill-posed problems, for which specific dedicated techniques have to be developed \citep[see][for example]{Tarantola, PijpersBook}. These methods require having an appropriate formalism and regularization of the problem, in line with the known or assumed properties of the physical problem under study. 

The inversion techniques used in asteroseismology are no exception to this rule. They most often derive from methods developed for helioseismology and for this reason have mostly been applied to solar-like oscillators, for which the underlying hypotheses of the inversion remain valid. It is worth mentioning that most seismic inversion techniques used in helio- and asteroseismology have actually been first applied to inferring the internal rotation of slowly rotating stars, including our Sun \citep[see][ and references therein]{Thompson2003}. Thanks to space-based photometry data, inferences of internal rotation have been made in various cases, from solar-like main-sequence stars \citep[see e.g.][]{Lund2014,Benomar2015,Schunker2016a,Schunker2016b,Benomar2018,Bazot2019}, to more massive oscillators \citep{Kurtz2014, Saio2015, Hatta2019,Hatta2022} and towards later evolutionary stages \citep[see e.g.][]{Deheuvels2014,Deheuvels2015,DiMauro2016,DiMauro2018,Deheuvels2020,Fellay2021} with great successes. 

In this short review, we will focus on seismic inversions of the internal structure of stars. We will start by discussing the goals of seismic inversions in Sect. \ref{Sec:Goals}. In Sect. \ref{Sec:Modelling}, we will discuss inversion techniques relying on the calibration of evolutionary models and mention examples of static approaches applied to specific cases in stellar evolution. In Sect. \ref{Sec:VariatEq}, we introduce the variational formalism applied in linear inversions, that will be discussed in Sect. \ref{Sec:LinearMethods} for both localized and indicator inversions. Finally, we will discuss non-linear inversions from the use of inner phaseshifts of the oscillations in Sect. \ref{Sec:NonLinear} and conclude in Sect. \ref{Sec:Conclusion}.

\section{Goals of seismic inversion techniques and underlying hypotheses} \label{Sec:Goals}

The goal of seismic inversion techniques is to provide a model of the internal structure that takes into account all the available observational constraints. This procedure can take various forms that we will discuss in the following sections.

From a mathematical point of view, the inversion procedure consists in solving the system of differential equations describing the stellar structure while taking all the available observational constraints into account. In practice, this already implies assuming a few hypotheses regarding the equilibrium state of the star and the physical processes considered in the description of its internal structure. 

The cases on which we focus will be slowly rotating solar-like stars without strong magnetic fields or being subject to tidal interactions capable of breaking the spherical symmetry. While these hypotheses might appear strong, they are in good agreement with observations and can be used for a wide range of targets. Hydrostatic and thermal equilibrium are also considered and the equilibrium equations of stellar structure are written in eulerian form as follows
\begin{align}
\frac{\partial m}{\partial r}&=4 \pi r^{2}\rho, \\
\frac{\partial P}{\partial r}&=\frac{-Gm\rho}{r^{2}}, \\
\frac{\partial l}{\partial r}&=4 \pi r^{2}\epsilon, \label{eqEnNew}\\
\frac{\partial T}{\partial r}& = \frac{-GmT\rho}{r^{2}P}\nabla, 
\end{align}
with $r$ the radial position in the spherically symmetric model, $m$ the mass of the sphere of radius $r$, $\rho$ the local density, $T$ the local temperature, $l$ the local luminosity, $P$ the local pressure,$\nabla=\frac{d \ln T}{d \ln P}$ the temperature gradient, and $\epsilon$ the local rate of energy generation. In addition to this system, the equation of state of the stellar material must be defined, namely the relation
\begin{align}
P=P(\rho,T,X_{j}),
\end{align}
with $X_{j}$ the chemical mixture of the stellar plasma. Alongside the determination of the internal structure, the inversion may also provide precise values of the fundamental stellar parameters such as mass, radius and age (when computing evolutionary models). In some specific cases, the full structure equations are solved in a static way, for example when the evolutionary path is unclear or difficult to compute with evolutionary models as for B-type subdwarf stars or white dwarfs for example. 

In other cases, the seismic inversion does not aim at determining the full structure. Its goal is rather to determine both the global parameters of the star such as mass and radius, the density profile inside the star, $\rho(r)$ and the first adiabatic exponent profile $\Gamma_{1}(r)=\frac{\partial \ln \rho}{\partial \ln P}\vert_{S}$.

In practice however, $\Gamma_{1}$ deviates only very little from $5/3$ in the stellar interior, apart from the ionization regions in the upper convective envelope. It is thus assumed constant or fixed at the value of a given reference structure used for the inversion. This situation is very different from helioseismology, where the quality of the data allows us to distinguish between various equations of state for the solar material based on the inversion of the $\Gamma_{1}$ profile in the convective envelope \citep[See e.g.][ and references therein]{Elliott1996,Elliott1998,Vorontsov2013}. 

Due to the lower quality of asteroseismic data and the absence of high-degree modes allowing to determine localized corrections in the upper layers, the equation of state of the stellar material will not be considered an issue\footnote{Constraints on the convective envelope can still be inferred from so-called ``acoustic glitches'', that can provide estimates of the helium abundance or the position of a sharp transition in temperature gradients at the bottom of the convective zone.}. This point will be further discussed in the following sections when presenting the variational formalism used in linear inversion techniques. 

In this context, the application of seismic inversions serve as a complement to evolutionary modelling. The inverted quantities can then be compared with those of evolutionary models and potentially constrain missing physical processes. Consequently, seismic inversions constitute an essential approach to improve the accuracy of stellar models and the determination of fundamental stellar parameters such as mass, radius and age. 

\section{Static and evolutionary modelling} \label{Sec:Modelling}

The inference of the internal structure of a star can take various forms. It may for example result from the adjustment of an evolutionary sequence to an observed target. For stellar systems such as binaries or clusters, simultaneous fitting of all members can even be done so that the modelling is more constraining, sometimes assuming the same chemical composition and age for all members. We hereby provide for the interested reader a few additional references on seismic modelling and asteroseismology of solar-like oscillators  \cite{RoxTools2002, JCD2010, Guzik2011,Goupil2011,Chaplin2013,DiMauro2016} as well as seismic studies of stellar clusters \citep{Handberg2017,McKeever2019} and binaries \citep[see e.g.][]{Bazot2016,Bazot2020,Salmon2021}. In other cases, an inversion can refer to the determination of the internal structure of a given star in a static way, departing from hypotheses regarding its evolutionary history. Such static approaches will sometimes even aim at determining only a key few internal quantities from dedicated formalisms while avoiding simplifying assumptions regarding the internal structure of the star. 

Such static approaches constitute powerful tools to further refine stellar evolutionary models and are often referred as ``seismic inversion'', while the use of evolutionary models is referred to as ``forward modelling''. This division is only present in the field of asteroseismology and can be misleading when studying inversion techniques used in other fields \citep[See][for example]{Tarantola, PijpersBook}. 

In what follows, we will briefly describe evolutionary and static inferences as applied to asteroseismic data. 

\subsection{Inferences from evolutionary models} \label{Sec:Evol}

The most common approach to determine the internal structure of a star is by coupling an optimization procedure to a stellar evolution code. While this is commonly referred as ``forward modelling'' in the seismic modelling community, it actually consists in a type of inversion in the mathematical sense of the word. 

In this type of inference, the structure of the star is coupled to an assumed evolutionary history. Namely, the evolution of the chemical abundances of the star is simulated by considering the effects of nuclear reactions and transport processes such as microscopic diffusion, macroscopic transport by rotation, convection, accretion and mass-loss. 

The main advantage of using evolutionary models resides in the possibility to test the evolutionary scenarios used to model stars and check the validity of the theory of stellar evolution. Moreover, in view of the recent needs of neighbouring fields such as exoplanetology and Galactic archaeology for fundamental stellar parameters such as masses, radii and ages, the use of optimization techniques on large grids of stellar models has now become routine, especially for main-sequence solar-like oscillators. Such modelling approaches have been applied to a wide range of targets from missions such as CoRoT, \textit{Kepler}, and TESS, as well as closer objects such as Alpha Centauri A$\&$B. 

In this context, numerous optimization techniques have been adapted to determine optimal stellar parameters. For example, local minimization techniques such as Levenberg-Marquardt algorithms have been applied in the past \citep[e.g.][]{Frandsen2002,Teixeira2003,Miglio2005}. However, due to the problem of local minima, especially with high-quality Kepler data, and the treatment of the uncertainties in such methods, global minimization techniques have been favoured such as Genetic algorithms, Markov Chain Monte Carlo approaches, or even Machine Learning software. Such methods have been applied to either precomputed grids of stellar evolutionary models or by computing the models on the fly to avoid relying on interpolation that could lead to a loss of accuracy. A few examples of such techniques and their associated results can be found in \cite{Mathur2012,Gruberbauer2013,Metcalfe2015,Bellinger2016,Rendle2019,Bazot2020,Silva2022}  .

A major drawback of the grid-based approach stems from the limitation to the physical ingredients of the precomputed models. Therefore, any change of abundance scale, opacity table, equation of state, or prescription for the mixing of chemical elements requires the whole grid to be recomputed, which can be time consuming. However, computing the models on the fly is not a viable option for sampling algorithms such as MCMC techniques that require large numbers of walkers to provide a proper distribution of the optimal parameters. 

\begin{figure}
\centering
  	\includegraphics[trim= 150 110 10 5 , clip, angle = 90, width=0.90\linewidth]{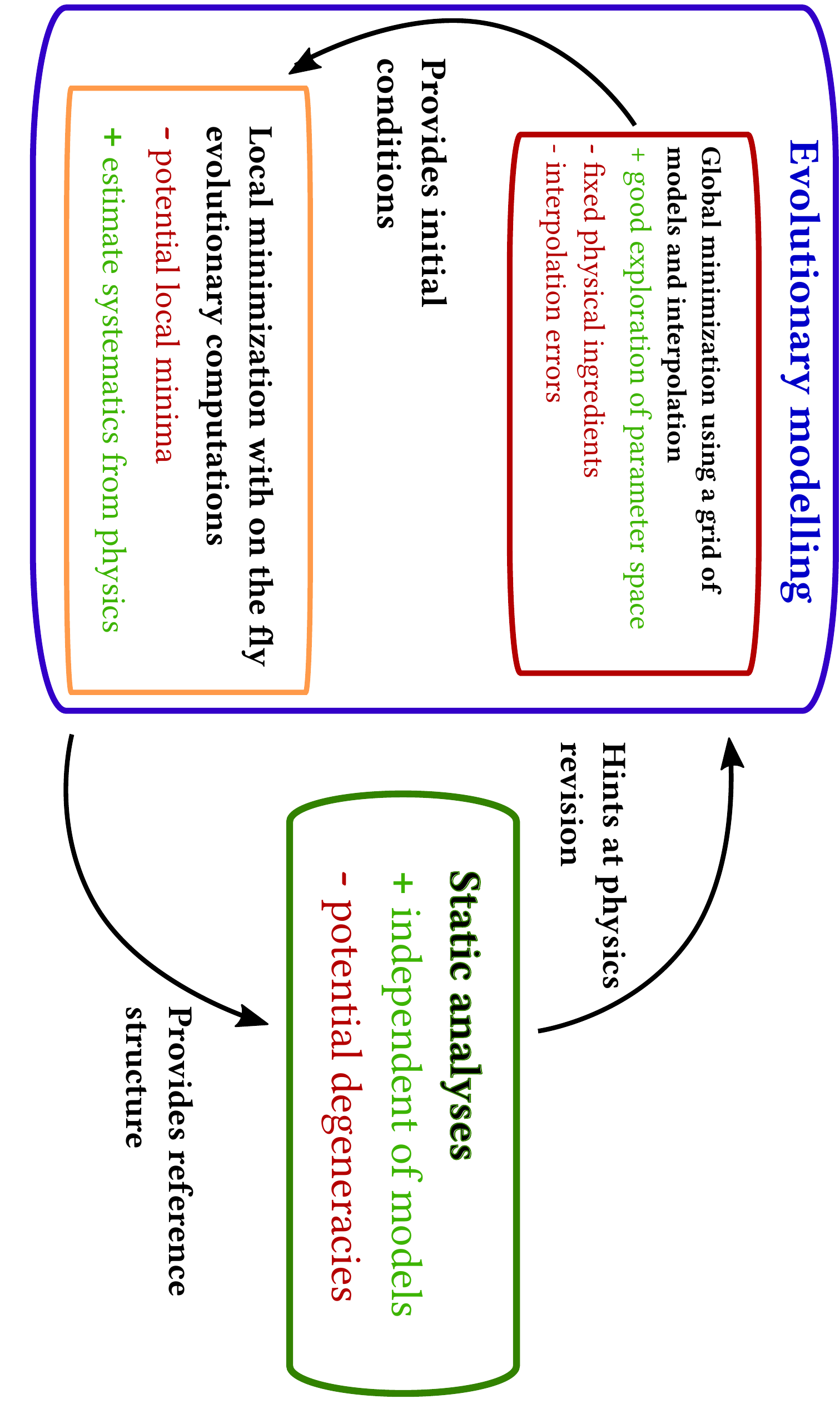}
  	\caption{Schematic representation of combinations of evolutionary and static inferences used in stellar modelling. Each box includes the strenghts (green) and weaknesses (red) of the various methods and the arrows represent their relations.}
		\label{FigSchema}
\end{figure}

More recently, a hybrid approach was presented \citep{Buldgen2019444,Betrisey2022}, where a combination of local and global minimization techniques are used. In the era of high-quality Kepler data, the uncertainties derived from solely propagating the observational error bars are small enough to sometimes be comparable, or smaller, than the changes in the optimal solution observed when varying the physical ingredients of the models. In this context, local minimization techniques offer the flexibility to estimate the change in the solution from a change of the physics, using the optimal solution obtained from an MCMC sampling of the parameter space. A combination of both is probably a good approach to test the impact of known physical processes on stellar fundamental parameters, while departing from an evolutionary history using static inferences might be optimal to study the presence of unknown physical processes. A schematic illustration of the interplay between the different modelling strategies is provided in Fig. \ref{FigSchema}.

Besides the algorithm used to determine the optimal model, the way seismic constraints are combined is also extremely important. One major issue of solar-like oscillations is their sensitivity to surface effects. To mitigate this issue, a first approach is to develop so-called empirical ``surface corrections'', that are based on solar frequencies \citep[See e.g.][]{RabelloParam,Kjeldsen2008,Ball2014} and/or on 3D averaged atmospheric models for which adiabatic oscillations are computed \citep[e.g.][]{Sonoi2015,Ball2016}. More recently, such analyses have been generalized to dipolar mixed modes by \cite{Ong2020, Ong2021}. 

\begin{figure}
\flushleft
  	\includegraphics[trim= 15 0 278 5 , clip, angle = 270, width=1.055\linewidth]{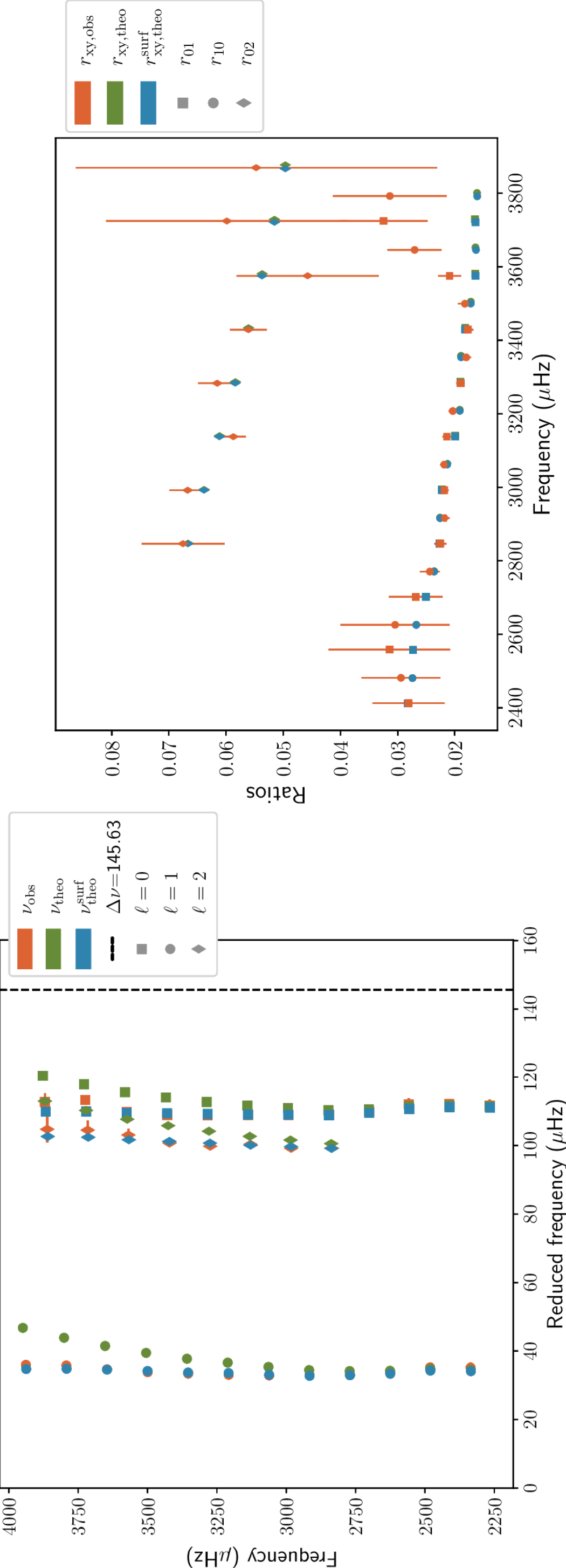}
		\caption{\textit{Left}: Illustration of the effects of surface corrections for the Echelle diagram of Kepler 93, showing the significant effect of the parametric corrections on high frequencies. \textit{Right}: Frequency separation ratios for the data of the left panel, showing their low sensitivity to the empirical surface corrections as the blue and reen symbols are essentially the same.}
		\label{FigEchelle}
\end{figure}

Given the amplitude of the surface effect, a direct fitting of the individual frequencies will be extremely sensitive to the empirical corrections. This is illustrated in the left panel of Fig. \ref{FigEchelle}, showing the individual frequencies of Kepler 93 and those of the associated optimal stellar model fitting them using the \cite{Ball2014} empirical surface correction. The actual amplitude of the surface correction is much larger than the uncertainties on the frequencies themselves, which may lead to strong biases, especially at high frequencies and can result in significant biases in the inferred stellar mass \citep{Jorgensen2020,Betrisey2022}.  

Another way to circumvent surface effects is to use combinations of frequencies, such as the so-called frequency separation ratios of the large and small frequency separations defined in \cite{RoxburghRatios} 
\begin{align}
r_{01}(n) &= \frac{\delta_{01}(n)}{\Delta \nu_1(n)}, \\
r_{10}(n) &= \frac{\delta_{10}(n)}{\Delta \nu_0(n+1)}, \\
r_{02}(n) &= \frac{\delta_{02}(n)}{\Delta \nu_1(n)},
\end{align}
with $\Delta \nu_l(n)$ the large separations, $\delta_{ij}(n)$ the small separations, and $n$ the radial order of the mode:
\begin{align}
\Delta \nu_l(n) &= \nu_{n,l}-\nu_{n-1,l}, \\
\delta_{01}(n) &=  \frac{1}{8}\left(\nu_{n-1,0}-4\nu_{n-1,1}+6\nu_{n,0}-4\nu_{n,1}+\nu_{n+1,0}\right), \\
\delta_{10}(n) &=  -\frac{1}{8}\left(\nu_{n-1,1}-4\nu_{n,0}+6\nu_{n,1}-4\nu_{n+1,0}+\nu_{n+1,1}\right), \\
\delta_{02}(n) &= \nu_{n,0}-\nu_{n-1,2}.
\end{align}

As shown in the right panel of Fig. \ref{FigEchelle}, these ratios are largely independent of the surface layers and thus much more efficient at constraining the internal structure of an observed target. Other techniques of surface-independent model fitting have been developed and presented in \cite{Roxburgh2015}. More recently, \cite{Farnir2019} also developed a comprehensive approach to decompose the spectrum of solar-like oscillations in seismic indicators uncorrelated to each other, taking into account both the smooth and the glitch component of the oscillation spectrum. 

In general, non-seismic constraints will also be considered when carrying out evolutionary modelling, such as the stellar luminosity, the effective temperature, the surface metallicity or the photospheric radius determined from interferometry (whenever available), or the mass if studying a spectroscopic binary system. While not always available, such non-seismic constraints, especially if determined with high precision, can prove extremely useful in lifting degeneracies between various solutions provided by pure seismic modelling.  

\subsection{Inferences from static models} \label{Sec:Static}

Static models are depictions of the internal structure of stars without considering their evolutionary history. In a strict sense, the observed oscillation frequencies carry information only on the current state of the star, but there can be multiple paths leading to the current observed properties. Static models are particularly useful to study stars in an evolutionary stage difficult to compute numerically, or for which the evolutionary path is unclear. 

Various techniques have been applied in the past to study the internal structure of SdB stars and white dwarfs \citep{Charpinet2008,VanGrootel2010,Giammichele2018,Charpinet2019,Fontaine2019}, providing constraints on the equation of state of dense stellar matter and the properties of semi-convective mixing in advanced evolutionary phases. 

Static seismic models have also been computed for the solar case  \citep[See e.g.][]{Basu1996,Takata1998,Shibahashi1999,Shibahashi2006,Buldgen2020} where the presence of high degree modes allows us to carry out a full scan via iterative methods. Some non-standard static models have also been computed by \cite{Hatta2021} for  KIC 11145123, a young $\gamma$-Doradus $\delta$-Scuti hydrid pulsator. As such, static modelling may also prove useful for modelling non-standard evolutionary products such as results of mergers, stripped cores, ...  for which pulsational properties may significantly differ from what their standard evolutionary counterparts would predict \citep[See e.g.][]{Deheuvels2022}. 

\section{Variational equations} \label{Sec:VariatEq}

The most commonly used inversion techniques rely on the so-called variational principle of adiabatic stellar oscillations \citep[See][]{Chandrasekhar1964, Clement1964,Chandrasekhar21964, LyndenBell1967}, that can be derived from the functional analysis of the adiabatic oscillation equations. It essentially states that at first order, perturbation of the adiabatic eigenfrequencies of gaseous spheres will be related to perturbations of the oscillation operator and link the seismic observables to interior quantities. In other words, small perturbations will follow, to first order, the following equation

\begin{equation}
\frac{\delta \nu^{n,\ell}}{\nu^{n,\ell}}=\frac{< \boldsymbol{\xi}^{n,\ell}, \delta \mathcal{L} (\boldsymbol{\xi}^{n,\ell}) >}{I^{n,\ell}},
\end{equation}

with $\nu^{n,\ell}$ the frequency of the oscillation mode, $\boldsymbol{\xi}^{n,\ell}$ its eigenfunction, $I^{n,\ell}$ the mode inertia \citep[see e.g.][for a general description of non-radial stellar oscillations]{Unno1989} and $\delta \mathcal{L}$ the perturbed operator of adiabatic oscillations. The notation $< . >$ denotes the scalar product over the functional space of the solutions of the adiabatic oscillation equations. It is defined as
\begin{align}
<\mathbf{a},\mathbf{b}> &= \int_{V} \mathbf{a} \cdot \mathbf{b} \rho dV,
\end{align}
with $V$ the volume of the sphere and $\rho$ the local value of the density in the sphere.

Examples of small perturbations in the stellar case include the effects of slow rotation, or minor mismatches between the internal structure of the model and the observed star. The former, that serves as initial condition for the inversion, will often be called ``reference model'', while the observed star or the target model will often be refereed to as the ``target'' of the inversion procedure. 

It is worth mentioning that while the variational principle is valid for ``small perturbations'' of an Hermitian operator (as is the general operator describing the full $4^{th}$ order system of adiabatic stellar oscillations), the domain of validity of the linear approximation is unclear, and likely depends on the oscillation modes and the quality of the reference model. The variational expressions have for example been generalized in the case of mixed oscillation modes by \cite{Ong2020} and \cite{Ong2021}. 

\begin{figure}
\centering
  	\includegraphics[trim=2 2 2 2, clip, width=0.90\linewidth]{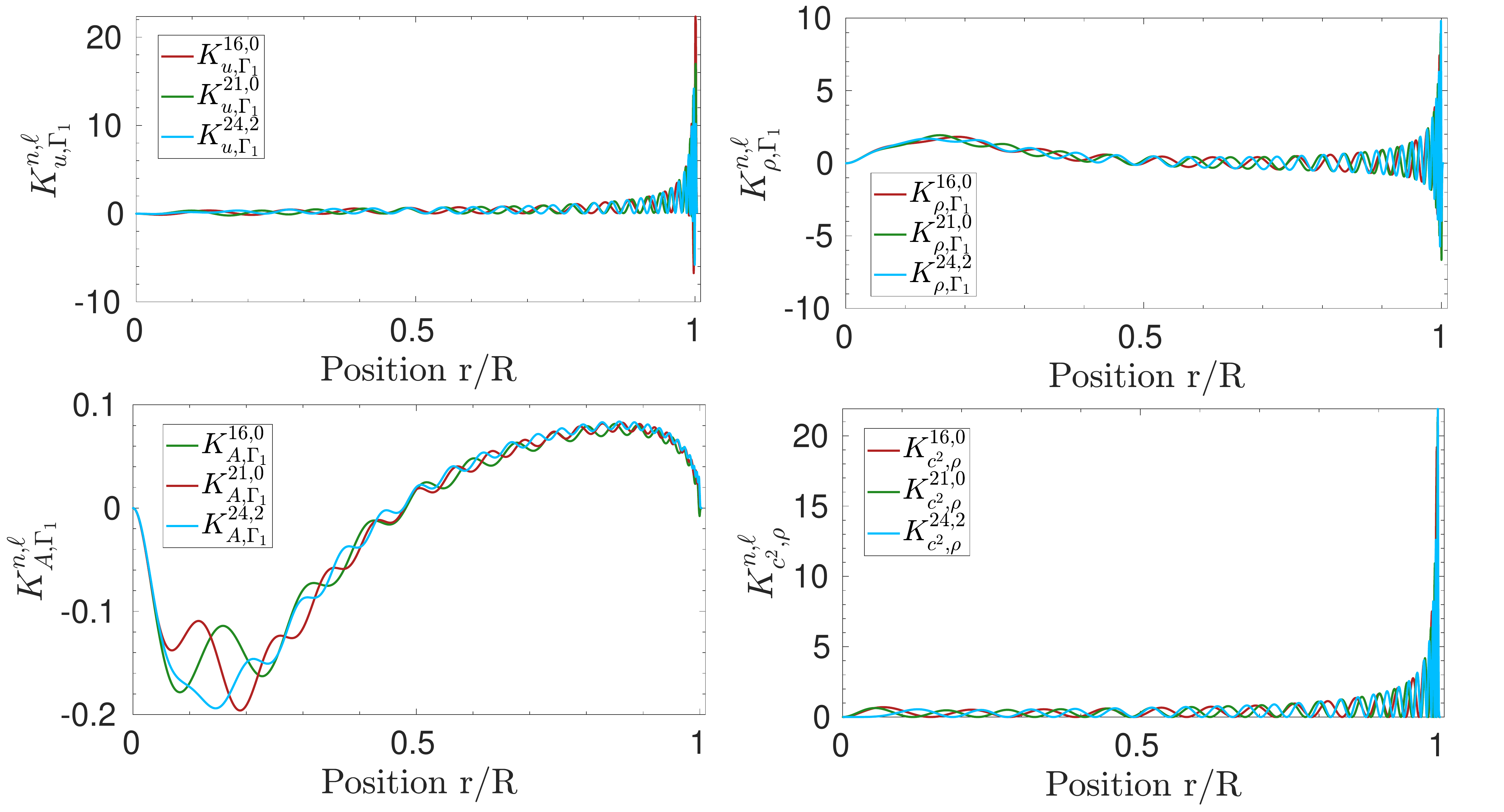}
		\caption{Structural kernels for various structural pairs and for low degree oscillations modes for a representative model of $16$CygA from \cite{Buldgen2022}. Each panel illustrates a different variable that can be used for variational inversions of the structure of a star as well as the degeneracy in the shape of the kernels for low degree modes.}
				\label{FigKerStruc}
\end{figure}

For the purpose of structure inversions, the equations are reworked to provide a formally simple expression allowing us to derive corrections to the internal structure of a given model \citep{Dziemboswki90}. This expression is 

\begin{align}
\frac{\delta \nu^{n,\ell}}{\nu^{n,\ell}}= \int_{0}^{R} K^{n,\ell}_{s_{1},s_{2}} \frac{\delta s_{1}}{s_{1}}dr + \int_{0}^{R}K^{n,\ell}_{s_{2},s_{1}} \frac{\delta s_{2}}{s_{2}}dr + \mathcal{O}(\delta^{2}), \label{eq:Variat}
\end{align}

where $s_{1}$ and $s_{2}$ are structural variables such as density, pressure, sound speed, ...  $K^{n,\ell}_{s_{j},s_{k}}$ the structural kernel related to variable $s_{j}$ in the $(s_{j},s_{k})$ pair which depends on the structure of the model and the eigenfunction of the oscillation mode. The notation $\delta$ defines a difference between the reference model and the observed target of a variable such as the frequency, the density as a function of radius, ... \citep[See][for the associated developments]{Dziemboswki90}. In Eq. \ref{eq:Variat}, the last term denotes that the formalism is valid to first order, and in some cases, higher order terms can become non-negligible, making the first order approximation inappropriate. Various pairs of structural kernels are illustrated in Fig. \ref{FigKerStruc},namely for the density, squared adiabatic sound speed, squared isothermal sound speed and Ledoux discriminant. The similarity between the kernels illustrates well the difficulties of asteroseismic inversions who will only have a few available modes to carry out the inversions.

The classical variational expressions are related to the adiabatic sound speed and density profiles. However, these expressions can be easily generalized to any physical quantity appearing in the adiabatic oscillation equations \citep{Elliott1996,Basu1997,Kosovichev2011,BuldgenKer}. In addition, assuming a linearized form of the equation of state can also be used to determine ``secondary'' variables such as temperature or in principle, chemical abundances. In such cases the $\Gamma_{1}$ perturbations are rewritten as

\begin{align}
\frac{\delta \Gamma_{1}}{\Gamma_{1}}=\frac{\partial \ln \Gamma_{1}}{\partial \ln P}\Big|_{\rho,Y,Z} \frac{\delta P}{P} + \frac{\partial \ln \Gamma_{1}}{\partial \ln \rho}\Big|_{P,Y,Z} \frac{\delta \rho}{\rho} + \frac{\partial \ln \Gamma_{1}}{\partial Y}\Big|_{P, \rho, Z} \delta Y + \frac{\partial \ln \Gamma_{1}}{\partial Z}\Big|_{P, \rho, Y} \delta Z,
\end{align}
with $P$, the local pressure, $\rho$, the local density, $Y$, the helium mass fraction and $Z$, the heavy element mass fraction. Other thermodynamic variables can be used in combination with the appropriate $\Gamma_{1}$ derivatives.

In practice however, this expression is not used in asteroseismology to carry out inversions of the chemical composition, but rather to naturally damp the contribution of the second integral in the inversion thanks to the low amplitude of the $\Gamma_{1}$ derivatives with respect to $Y$. Such ``tricks'' lead to more stable inversions for which the variational equation almost reduces to an integral expression with one structural variable instead of two. We will come back to this point in Sect. \ref{Sec:LinearMethods}.

Lagrangian perturbations can also be considered in Equation \ref{eq:Variat}. In this case, the perturbations of the structural variables will be considered at fixed mass instead of fixed radius \citep{JCD1997}. Instead of using individual frequency differences, it is also possible to express the variational equations for frequency separation ratios, as was shown in \cite{Oti2005} and applied in \cite{Betrisey2022B}.

One major weakness of the variational expressions using individual frequencies is their strong sensitivity to surface effects. They indeed rely on the adiabatic approximation, which is not valid in the upper layers where the thermal and dynamical timescales are of the same order of magnitude. Moreover, adiabatic oscillation codes often use simplified boundary conditions and the poor modelling of the upper convective layers by the mixing-length theory will lead to inaccuracies of the oscillation frequencies at a significant level with respect to the observational errors. Such sensitivity can however be damped when using frequency separation ratios for the inversion. 

Indeed, as they directly use frequency differences, the variational equations will suffer to some extent from the same caveats as directly using the individual frequencies as constraints in evolutionary or static modelling. This implies that the variational expression \ref{eq:Variat} will have to be supplemented by a surface correction term that often takes a polynomial form

\begin{align}
\mathcal{F}(\nu)=\sum_{k}c_{k}(\nu)\nu^{k}, \label{eq:SurfCorr}
\end{align}

with $\nu$ a given frequency and $c_{k}$ the surface correction coefficients associated with the power $k$ of the frequency. Existing surface corrections include the classical polynomial approach of degree $7$ applied in helioseismology\citep{RabelloParam}, the two-terms \cite{Ball2014} correction and the \cite{Sonoi2015} formula that can also be linearized, or applied to the frequencies beforehand using their empirical relation of the surface correction with effective temperature and surface gravity.   

It is therefore important to keep in mind that the validity of the variational relations is limited. It can for example be measured using error functions such as 

\begin{align}
\mathcal{E}^{n,\ell}=\frac{\Lambda^{n,\ell}_{\mathrm{LHS}} - \Lambda^{n,\ell}_{\mathrm{RHS}}}{\Lambda^{n,\ell}_{\mathrm{LHS}}}, \label{Eq:ErrorVar}
\end{align}

with $\Lambda^{n,\ell}_{\mathrm{LHS}}$ the relative frequency difference and $\Lambda^{n,\ell}_{\mathrm{RHS}}$ the right-hand side of Eq. \ref{eq:Variat}. 

\begin{figure}
\centering
  	\includegraphics[trim=2 2 2 2, clip, width=0.95\linewidth]{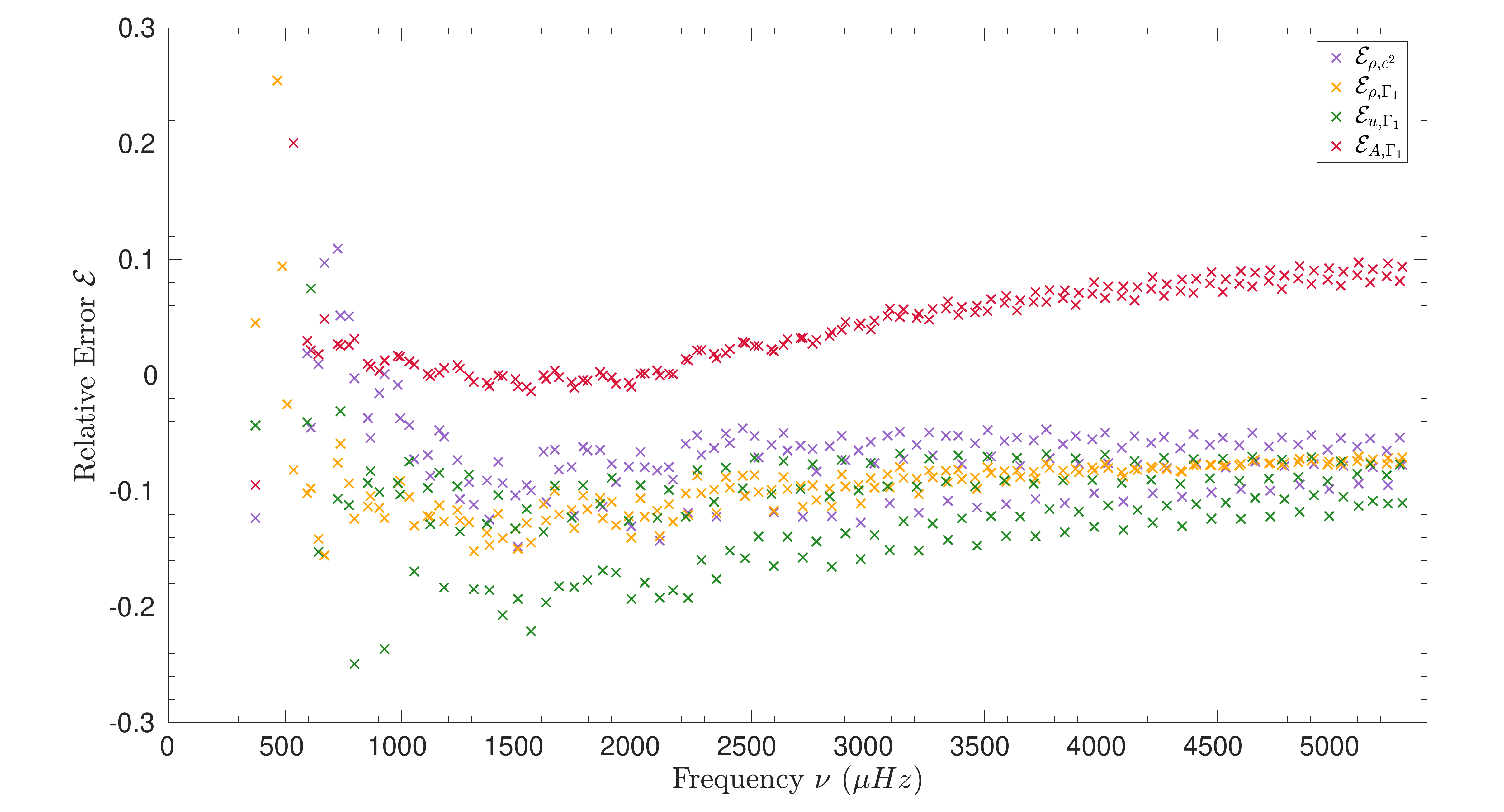}
     \caption{Illustration of the verification of the integral relation between relative frequency differences and relative structure differences for solar models including various transports of the mixing of chemicals, using Eq. \ref{Eq:ErrorVar}, using models from \cite{BuldgenKer} and various structural pairs, namely $(\rho,c^{2})$, $(\rho,\Gamma_{1})$, $(u,\Gamma_{1})$ and $(A,\Gamma_{1})$.} 	
		\label{FigVarError}
\end{figure}

As illustrated in Fig. \ref{FigVarError}, the accuracy of the variational inversion will vary depending on the oscillation modes considered as well as the structural variables used. Such considerations might be very important when choosing the most adapted variable pair for a given target and set of observed frequencies before carrying out an inversion of the structure. 

\section{Linear inversion techniques - the SOLA Method} \label{Sec:LinearMethods}

The variational equations provide the basis for the use of linear inversion techniques in helioseismology. Namely, under the hypothesis of the validity of the linear integral relation \ref{eq:Variat}, these equations can be solved to provide corrections to structural variables. The linear approaches provide one step of correction, and are not iterated, unlike the non-linear methods we present in Sec. \ref{Sec:NonLinear}. As mentioned above, the variational equations can be rewritten for a wide range of physical quantities, offering some degree of freedom regarding the target of the inversion. 

However, the situation is in practice far more complex. Linear inversion techniques such as the OLA \citep{Backus1970}, SOLA \citep{Pijpers} or RLS methods \citep{Tikhonov1963} have been adapted for inverting helioseismic data \citep[see][for a comparison between OLA and RLS]{JCD1990,Sekii1997} and \citep[see][for a review]{Reese2018}. Inversions of stellar structure using asteroseismic data have so far been limited to the use of the SOLA method\footnote{We note however that instances of RLS inversions of the internal rotation of distant stars can be found in \cite{Deheuvels2014,Schunker2016}}, which is the one we will describe here. 

The philosophy behind the SOLA inversion is to compute linear combinations of the relative frequency differences based the following cost function

\begin{align}
\mathcal{P}(c_{i})=&\int_{0}^{R}\left[ K_{\rm{avg}}(r)-\mathcal{T}(r)\right] dr + \beta \int_{0}^{R}K^{2}_{\rm{cross}}(r)dr \nonumber \\ & +\lambda \mathcal{N}+\tan \theta \frac{\sum^{N}_{i}(c_{i}\sigma_{i})^{2}}{<\sigma^{2}>}+\sum^{N}_{i}c_{i}\sum^{L}_{k}a_{k}\psi_{k}(\nu_{i}), \label{eq:CostSOLA}
\end{align}
with $\mathcal{T}$ the target function of the inversion, $\lambda$ a Lagrange multiplier, $\mathcal{N}$ an additional regularization term (as will be discussed later), $c_{i}$ the inversion coefficients, $\theta$ and $\beta$ the trade-off parameters, $\sigma_{i}$ the uncertainties of the relative frequency differences, $<\sigma^{2}>=\frac{1}{N}\sum_{i=1}^{N}\sigma^{2}_{i}$ and $N$ the number of observed frequencies. The term $\sum^{L}_{k}a_{k}\psi_{k}(\nu_{i})$ is a polynomial expression for the surface correction defined in Eq. \ref{eq:SurfCorr}, with $L$ the number of surface terms in the polynomial definition. In addition to these quantities, we define in Eq. \ref{eq:CostSOLA} two terms, $K_{\rm{avg}}$ and $K_{\rm{cross}}$, the averaging and cross-term kernels, defined from the recombination of the structural kernels with the inversion coefficients. They will have the form
\begin{align}
K_{\rm{avg}}&=\sum_{i}^{N}c_{i}K^{i}_{s_{1},s_{2}}, \label{eq:AVGKer}\\
K_{\rm{cross}}&=\sum_{i}^{N}c_{i}K^{i}_{s_{2},s_{1}}, \label{eq:CrossKer}
\end{align}
if the target function of the inversion is related to the variable $s_{1}$ in the $(s_{1},s_{2})$ structural pair. 

In other words, the SOLA method is based on a trade-off between precision and accuracy. The goal is to compute the best fit to the target function of the inversion while avoiding to amplify too much the observational uncertainties. Looking at Eq. \ref{eq:AVGKer}, it is straightforward to see that the number of observed frequencies will play a crucial role for the accuracy of the method. The more frequencies are observed, the more structural kernels are available for recombination, and the more accurate the inversion will be. 

The situation in asteroseismology is however very far from that of helioseismology. Since the surface of the star is not resolved, geometric cancellation forbids the detection of solar-like oscillations of angular degree higher than $3$. This has important consequences for the capabilities of linear inversions, as the datasets for the best \textit{Kepler} targets count at most $\approx 50$ to $\approx 60$ individual frequencies with low degrees. In such conditions, full scans of the internal structure, as carried out for the Sun, are not achievable. 

Most of the time, the linear inversion of the structure will be limited by the number of frequencies and the validity of the linear formalism. An important difference between asteroseismic and helioseismic inversions is that the fundamental parameters, mass, radius and age of the star under study are not known. The solar case thus consists in an ideal environment where the bounds of the integrals in Eq. \ref{eq:Variat} are known, and where the knowledge of the age of the Sun provides a degree of control on the computed models, ensuring the validity of the linear formalism. 

\begin{figure}
\centering
  	\includegraphics[trim=2 2 2 2, clip, width=0.95\linewidth]{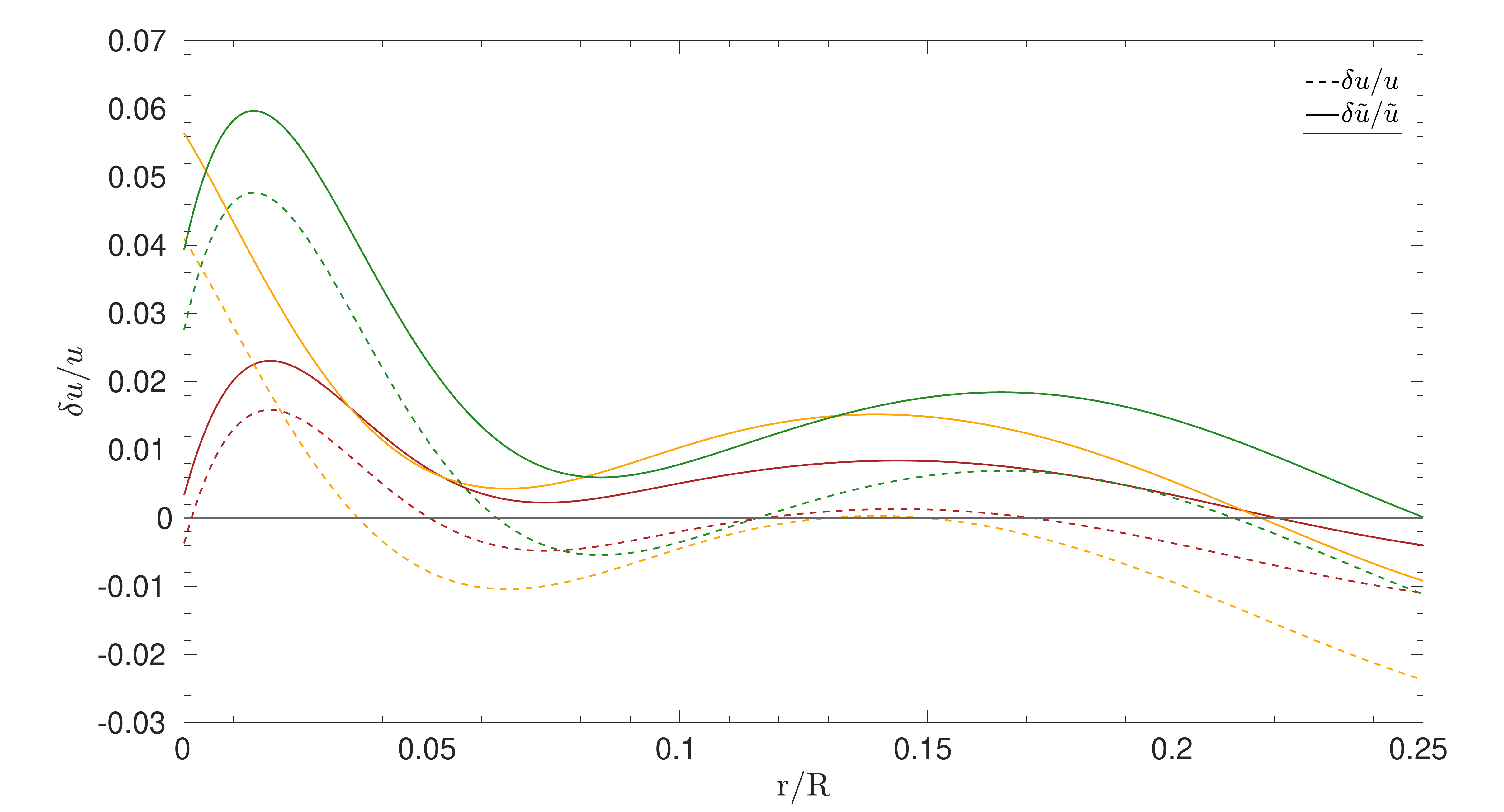}
     \caption{Differences between the dimensional ($(u)$, dashed line) and dimensionless ($\tilde{u}$, plain line) squared isothermal sound speed for models of $16$CygA from \cite{Farnir2020}. The differences are those computed in \cite{Buldgen2022}, Fig. A.13 and are linked to models of $16$CygA showing differences in their mean density values between $1$ (orange and green lines) and $2\%$ (red lines).} 	
		\label{FigDimAdimU}
\end{figure}

This will not be the case in asteroseismology, implying that a workaround has to be found, and some care has to be taken when interpreting the results of linear inversions. In practice, one can include an additional term in the cost function of the inversion related to the minimization of the mean density, as suggested by \cite{Roxburgh98}. In such condtions, one can consider that using dimensional and dimensionless frequencies, denoted $\nu$ and $\tilde{\nu}$ would be equivalent and that
\begin{align}
\frac{\nu_{\mathrm{obs}}-\nu_{\mathrm{ref}}}{\nu_{\mathrm{ref}}}=\frac{\tilde{\nu}_{\mathrm{obs}}-\tilde{\nu}_{\mathrm{ref}}}{\tilde{\nu}_{\mathrm{ref}}}, \label{eq:DimAdim}
\end{align}
if the mean density is perfectly fitted. This implies that the actual corrections derived by the inversion are for the dimensionless variable and not its dimensional counterpart. Another pragmatic approach to take this scaling effect into account is to carry out inversion for ensembles of models with the same mean density, determined for example from mean density inversions (See Sect. \ref{Sec:Indic}). This ``scaling'' effect can become very important when comparing actual differences between models. It also illustrates some degree of degeneracy in seismic inferences, and shows the importance of independent radii estimates from interferometric measurements, or combination of parallaxes and spectroscopic measurements. In practice, any high-quality non-seismic constraints will prove very useful to seismic modelling, and also explains why binary systems are still key testbeds of stellar physics, even in the era of space-based photometry missions \citep{Metcalfe2015,Appourchaux2015,Bazot2016,Farnir2020,Salmon2021}. Figure \ref{FigDimAdimU} illustrates such differences for two models of the \textit{Kepler} target $16$CygA from \cite{Farnir2020}. The behaviour of the dimensional and dimensionless relative differences in squared isothermal sound speed can be quite different, showing the importance of understanding what variable is actually constrained by the inversion. 

Another limitation of linear inversions is the range of validity of the linear formalism. In practice, calibrations using only non-seismic parameters will not be sufficient to ensure the applicability of the variational formalism to the inverse problem. Thus, as already noted by \cite{Thompson2002}, a preliminary form of seismic modelling must be carried out before the inversion is performed, which implies that seismic data is already used beforehand. From a modelling point of view, this means that the inversion step is not completely independent of the preliminary modelling procedure, and that inversions relying on the variational equations can often be biased. This problem can become more substantial if the amplitification of the observational uncertainties is small, meaning that the inversion appears artificially precise. In this context, providing a set of reference models large enough is very important to ensure that the precision of the inversion is correctly assessed from the point of view of model-dependencies, non-linearities, and impact of empirical surface corrections. 

\subsection{Dealing with surface effects}

The additional surface correction term in Eq. \ref{eq:SurfCorr} leads to various complications in asteroseismic inversions. The additional term in Eq. \ref{eq:CostSOLA} can often have a very significant impact on the inversion results, as it leads to a less favourable trade-off for a method already limited by the low number of observed modes to fit the target function. Determining the actual impact of the surface corrections on inversion results is particularly important, as in some cases they will lead to a larger variation of the inverted result than the uncertainties derived from the SOLA method. This is the case for mean density inversions, which thus require a more careful analysis. Indeed, the precision on the determined mean density will have an impact on the determined stellar mass using evolutionary modelling or the determined planetary mass from a radial velocity curve.

The classical polynomial fit used for helioseismic inversions is not applicable in the context of asteroseismology. It was usually advised to consider a polynomial of order up to $7$ to fit the ``surface term'' \citep{RabelloParam}, which is not possible with asteroseismic data. \cite{Reese2012} initially attempted to limit the correction to the first order, but tests in hare and hounds exercises showed that this was even less efficient than not considering any correction \citep{Reese2016}. 

With the advent of the new empirical corrections of \cite{Ball2014} and \cite{Sonoi2015}, the surface term could be reduced to two additional terms. In such conditions, the surface term can sometimes be directly included in the fitting of the SOLA cost-function with limited impact on the quality of the fit of the target function of the inversion regarding accuracy, but will actually impact the amplitude of the inversion coefficients and thus the precision of the inversion. However, with very limited datasets, or for some more unstable inversions, the addition of two terms in the cost-function can lead to low quality reproduction of the target. 

In such conditions, surface corrections can also be applied following the empirical formula of \cite{Sonoi2015} as a function effective temperature and surface gravity before carrying the inversion, using the corrected frequencies as the observed ones. An analysis of the importance of surface corrections for mean density inversions on red giant stars using averaged 3D atmospheric models and non-adiabatic frequency computations has been carried out in \cite{Buldgen2019}. They showed that applying the surface corrections beforehand could be more efficient than including it in the cost-function, as it did not affect the fit of the target function. A similar observation was made for Kepler 93 by \cite{Betrisey2022}. However, further tests and more thorough analyses have to be carried out for other indicators and datasets before concluding on the best approach to take into account surface effects. In addition, such studies do not necessarily indicate that the surface corrections are ultimately accurate, and an efficient workaround is then to avoid the surface effect dependency in the inversion altogether, as will be discussed in Sect. \ref{Sec:NonLinear}. 

\subsection{Inversions of localized corrections} \label{Sec:Local}

The original SOLA inversions have been designed for determining local average corrections of the solar rotation profile \citep{Pijpers} from the variational expressions applied in the slowly rotating case. The same approach can be directly applied to structural inversions, with the only major change being that a cross-term contribution appears due to Eq. \ref{eq:Variat} including two integrals compared with the single integral relation of rotation perturbations. 

Inversions of localized averages have been extensively performed in helioseismology, as well as comparisons with the RLS technique and applications to various structural pairs. The application of SOLA inversions to asteroseismic data was foreseen quite early, with studies on artificial data already carried out in the 1990s and early 2000s \citep{Gough1993,Gough932,Roxburgh98,Basu2002,Takata2002}. In most cases, the expected quality of the dataset was actually higher than what was actually brought by the space-based photometry missions, with some artificial datasets going as high as $100$ observed frequencies, which unfortunately has not been achieved for any solar-like oscillator observed by \textit{Kepler}.

In the case of localized inversions, the target function of the inversion usually is a Gaussian-like function of the form
\begin{align}
\mathcal{T}=\alpha r \exp\left(-\left( \frac{r-y}{\Delta} + \frac{\Delta}{2r_{0}} \right)^{2} \right),
\end{align}
with $\alpha$ a normalisation constant, $y$ the center of the Gaussian target function  and $\Delta=\frac{\Delta_{A}c(y)}{c_{A}}$ is linked to the width of the Gaussian, $\Delta_{A}$ being a free parameter, and $c(y)$ and $c_{A}$ the adiabatic sound speed at the maximum of the kernel and at a radius of $0.2R$, respectively.

The regularisation term $\mathcal{N}$ of Eq. \ref{eq:CostSOLA} is then a unimodularity constraint on the averaging kernel of the form
\begin{align}
\int_{0}^{R}K_{\rm{avg}}dr=1,
\end{align}
to avoid high amplitudes of the inversion coefficients that can make the procedure unstable. 

\begin{figure}
\centering
  	\includegraphics[trim=2 2 2 2, clip, width=0.95\linewidth]{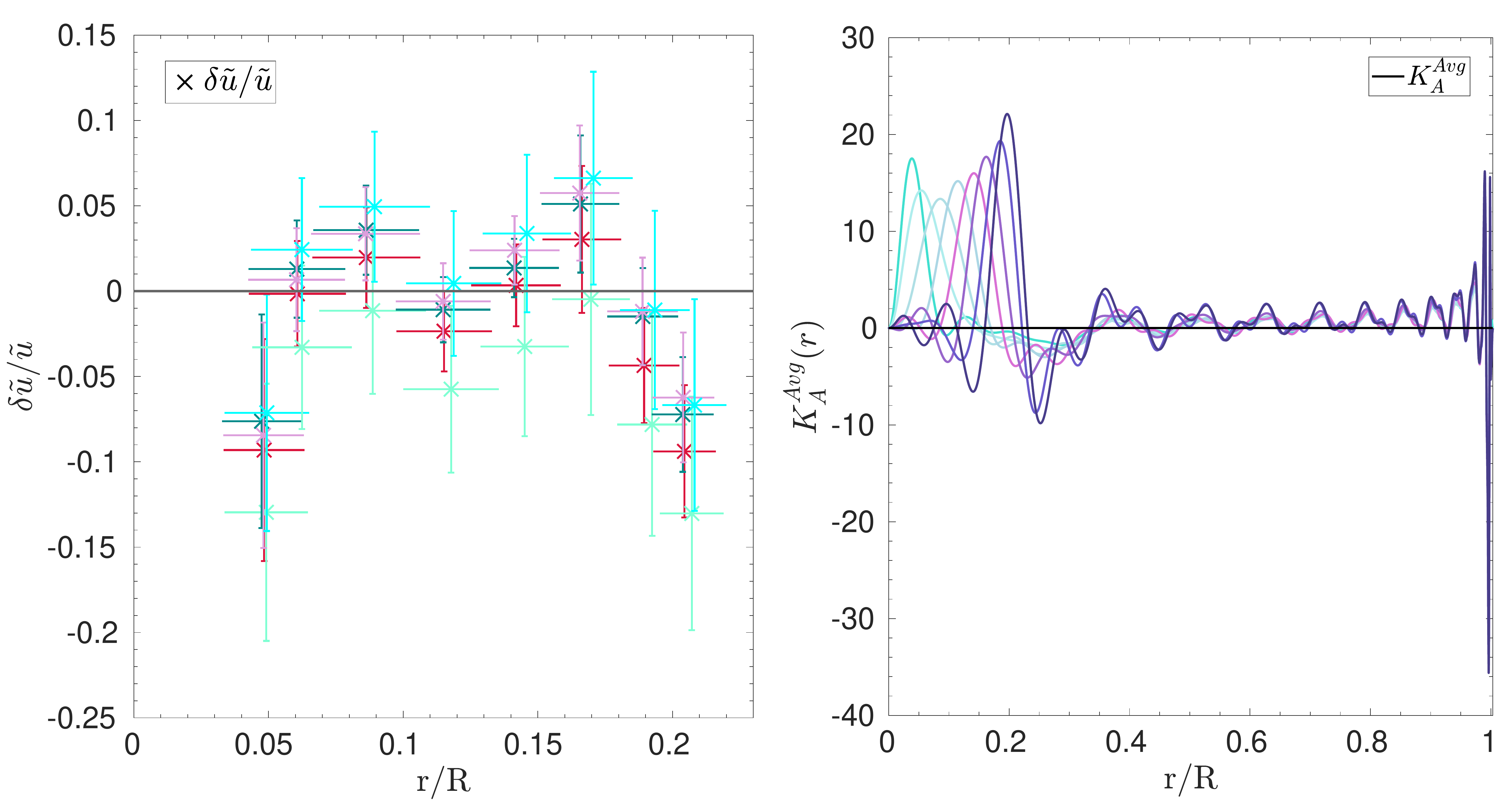}
  	\caption{\textit{Left}: relative differences in dimensionless squared isothermal sound speed for $16$CygA as a function of normalized radius. Each colour represents a different reference model for the inversion. The set of reference models is that of \cite{Buldgen2022}, Table 2. \textit{Right}: averaging kernels for the localized inversion of 16CygA using the $(u,Y)$ structural pairs (adapted from \cite{Buldgen2022}).}
		\label{FigUInv}
\end{figure}

Applications of the SOLA method in its original form to actual asteroseismic data can be found as early as $2004$ \citep{DiMauro2004}. More recent applications aiming at determining localized corrections to \textit{Kepler} targets can be found in \citep{Bellinger2017, Bellinger2019, Kosovichev2020, Bellinger2021, Buldgen2022}. Such applications have been limited to the best \textit{Kepler} targets, such as the 16Cyg binary system. Even in such cases, localised kernels can only be obtained in the deep layers, as a result of the availability of only low degree modes. An illustration of Gaussian averaging kernels obtained for 16CygA is illustrated in the right panel of Fig. \ref{FigUInv}.

As shown in the left panel of Fig \ref{FigUInv}, such inversions have confirmed the accuracy of the reference evolutionary models for both stars determined from asteroseismic modelling using the method of \cite{Farnir2020}. The fact that the SOLA inversion cannot pinpoint differences between models of solar twins is not really a surprise. Indeed, the effects of varying the transport of chemicals by inhibiting settling of heavy elements to mimick turbulence at the base of the convective envelope, changing the radiative opacity tables, or changing the reference solar abundances is rather small for a given set of parameters such as mass, radius and age. From the analysis of solar models, the differences seen are of the order of $1\%$ at the base of convective envelope and reduce to $0.2\%$ in the deep radiative layers. In these conditions, linear asteroseismic inversions, showing uncertainties of a few percent, might not be able to pintpoint such small differences in structure if the fundamental parameters of a star are well constrained. 

This points towards a quite stringent restriction of linear inversions, as they have a sort of ``niche'' where their application might be useful. In practice, an unsatisfactory agreement between the observed and modelled frequency separation ratios, $r_{02}$ or $r_{01}$, may indicate that carrying out inversions of the internal structure can be useful, as shown in \cite{Bellinger2019}. However, if all frequency separation ratios are well reproduced, linear inversions of the internal structure might only confirm the validity of the evolutionary models and the traces of additional processes such as macroscopic mixing of chemicals might be looked for using frequency glitches \citep{Monteiro2005, Mazumdar2012,Mazumdar2014,Verma2014,Verma2017,Verma2019a,Verma2019b} or depletion of light elements such as Lithium and Beryllium \citep{Deal2015}.

Localised inversions have also been applied to post main-sequence \textit{Kepler} targets such as in \cite{Kosovichev2020} and \cite{Bellinger2021}. However, the validity of the variational equations in the context of mixed modes still needs to be thoroughly investigated, as shown by \cite{Ong2020} and \cite{Ong2021}, as coupling effects can cause the classical relation to break down for mixed modes, implying that a non-linear formalism, taking properly into account the coupling using the full system of oscillation equations might be required. In this context, non-linear inversions appear the most favourable approach for extracting meaningful constraints from asteroseismic data. 

\subsection{Inversions of global indicators} \label{Sec:Indic}

As a result of the difficulties of carrying out full profile inversions, a compromise was struck by \cite{Reese2012} who decided to focus on global quantities rather than attempt at localizing kernels. The main goal of such ``indicator'' inversions is to focus on one single quantity at a time using the variational expressions, trying to extract constraints on some well-chosen key quantities instead of detailed profiles. The chosen quantities are determined based on the structural properties under investigation, such as the mean molecular weight gradient in the deep radiative layers, or the profile of an entropy proxy close to the border of a convective zone. 

A main difficulty of linear asteroseismic inversions is to find appropriate target functions for SOLA inversions that can be easily fitted with a very limited number of frequencies. Consequently, \cite{Reese2012} focused on a physical quantity well-known to be constrained by solar-like oscillations, the mean density. \cite{Buldgen2015tau,Buldgen2015, Buldgen2018} then focused on generalizing the formalism to other physical quantities. The linear inversion of an indicator will be computed as follows
\begin{align}
\frac{\delta E}{E} = \sum^{N}_{i}c_{i}\frac{\delta \nu_{i}}{\nu_{i}}, \label{eq:linearIndicCorr}
\end{align} 
with $E$ the indicator, $c_{i}$ the inversion coefficients determined by the SOLA method and $\nu_{i}$ the individual frequencies. 

The indicator is related to structural variables using an integral definition 
\begin{align}
E=\int_{0}^{R}f(r)g(s_{1})dr, 
\end{align}
and applying an eulerian linear perturbation to this equation leads to 
\begin{align}
\frac{\delta E}{E}=\int_{0}^{R}\frac{s_{1}f(r)}{E}\frac{\partial g(s_{1})}{\partial s_{1}}\frac{\delta s_{1}}{s_{1}}dr = \int_{0}^{R}\mathcal{T}_{E}\frac{\delta s_{1}}{s_{1}}dr,
\end{align}
which defines the target function of the indicator $\mathcal{T}_{E}$ which will be used in Eq. \ref{eq:CostSOLA} for this specific inversion. 

An additional term is introduced in the SOLA cost-function, denoted $\mathcal{N}$ in Eq. \ref{eq:CostSOLA}. For the case of indicator inversion, it takes the following form
\begin{align}
\mathcal{N}=\left[ k - \sum^{N}_{i}c_{i}\right], \label{eq:RegTermIndic}
\end{align}
with $k$ an integer that relates the scaling of the indicator with respect to mass. The argument is based on the relation between an indicator and the frequencies, which go as $\sqrt{\frac{GM}{R^{3}}}$. Since the radius is essentially fixed by the definition of the boundary of the integral variational relation, the scaling ends up being a scaling with mass. The idea is to determine the properties of a homologous transformation that will lead to the correct rescaling of the indicator value. 

Essentially, if the indicator goes as $E\propto M^{k/2}$, then a small perturbation of $\delta E/E = \epsilon$ will lead to a small perturbation of the frequencies $\delta \nu/\nu = \epsilon/k$. In these conditions, it can be shown that a homologous transformation with $\sum_{i}c_{i}=k$ leads to the correct rescaling of the indicator. Such inversions have been called ``unbiased'' in \cite{Reese2012}. In practice, this term acts as an additional regularization term for the inversion to avoid high amplitudes of the inversion coefficients. As seen above, similar regularization terms are introduced for localized inversions in the form of unimodularity constraints for the kernels. 

This additional constraint has also been associated with simple ``non-linear'' generalizations of the indicator inversions, following the approach of an iterative scaling of the model using homologous relations. This iterative scaling can be shown to provide generalized formulations of Eq. \ref{eq:linearIndicCorr} which will depend on the value of the additional coefficient in the regularization term. 

\begin{figure}
\centering
  	\includegraphics[trim=2 2 2 2, clip, width=0.65\linewidth]{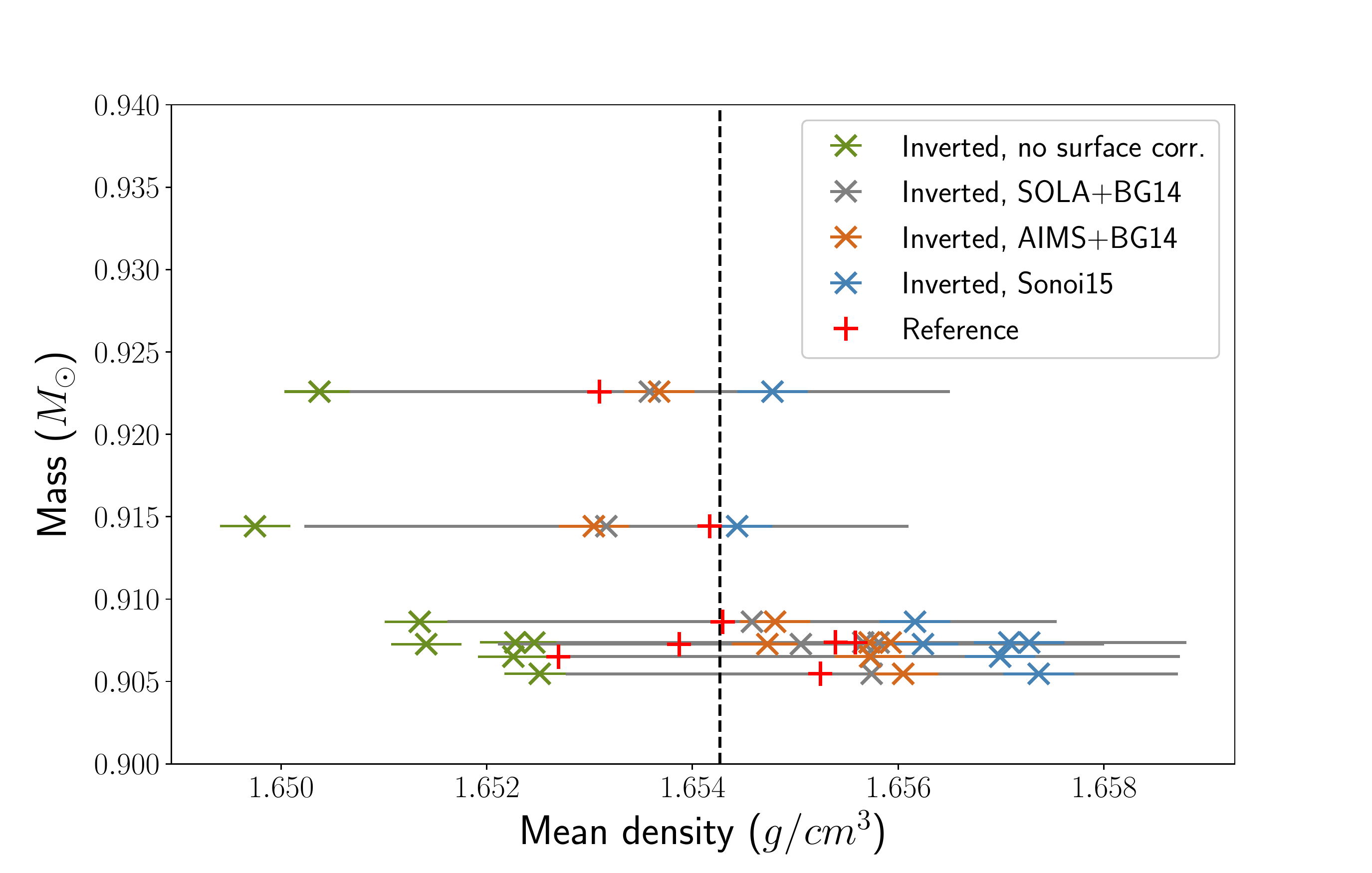}
  	\caption{Effect of surface corrections and model-dependency on mean density inversions of Kepler 93 (adapted from \cite{Betrisey2022}) as a function of mass of the reference model. Red crosses show the reference mean density values, olive green crosses show results without surface corrections, blue and grey crosses show results using the \cite{Sonoi2015} and \cite{Ball2014} corrections in the SOLA cost function, whereas brown crosses show results using coefficients extracted from MCMC modelling and applied beforehand.}
		\label{FigRhoKep93}
\end{figure} 

\subsubsection{Mean density}

Mean density inversions were defined in \cite{Reese2012}, as a potential application of the SOLA method to refine the determinations of stellar fundamental parameters from asteroseismic data. These inversions have been further tested later on in \cite{Buldgen2015tau} for solar like stars, and applied to radial oscillations of post-main sequence stars in \cite{Buldgen2019}. The integral definition of the stellar mean density is
\begin{align}
\bar{\rho}=\frac{3}{4\pi R^{3}}\int_{0}^{R}4\pi r^{2}\rho dr.
\end{align}

Hence, the target function is given by
\begin{align}
\mathcal{T}_{\bar{\rho}}(r)=\frac{4\pi\rho r^{2}}{R^{3}\rho_{R}}, \label{eq:Targetrho}
\end{align}
with $\rho_{R}=\frac{M}{R^{3}}$. An illustration of the target function in the case of 16CygA is provided in the bottom right panel of Fig \ref{FigTarIndic}. This function is usually fitted using structural kernels of the $(\rho, \Gamma_{1})$ structural pair in the variational equation.

Mean density inversions have been applied to a wide variety of targets, mostly on the main sequence. They offer the advantage of providing an accurate determination of the mean density, beyond the capabilities of asymptotic estimates such as the large frequency separation \citep{Vandakurov1967} and being less sensitive to surface effects than the latter. 

Due to the easily fitted target function, mean density inversions are one of the few that can be applied with a very limited number of observed frequencies \citep[See][]{Reese2012,Buldgen2015tau}, without recalibration of the trade-off parameters. These inversions are thus suitable for an ``automated' approach in modelling pipelines. However, they suffer from an overestimated precision of the inversion. As a result of the shape of the target function, the inversion coefficients are not very large and the amplification of the uncertainties is small. While this may seem as an advantage, it also means that the error bars on the determined mean density cannot be directly estimated using the SOLA method. In fact, the spread of results observed when using multiple models and empirical surface corrections is often much larger than the $1\sigma$ error bars provided by the SOLA method. This is illustrated in Fig. \ref{FigRhoKep93} for Kepler 93. In practice, mean density inversions must then be applied to a given set of models and their precision is closer to $0.2\%$ or $0.3\%$ \citep[See][]{Betrisey2022} rather than the claimed $0.1\%$ or less computed from the propagation of the observational uncertainties. 

\subsubsection{Acoustic radius}

Acoustic radius inversions were defined in \cite{Buldgen2015tau}. The acoustic radius of a star is defined as
\begin{align}
\tau = & \int_{0}^{R}\frac{dr}{c} \label{eqacousticradius},\\
\end{align}
and is related to the large frequency separation in the asymptotic regime as $\Delta \nu = 2/\tau$. 

These inversions can be carried out with both the $(c^{2}, \rho)$ or the $(\rho,\Gamma_{1})$ structural pair. As discussed above, structural pairs involving either $\Gamma_{1}$ or $Y$ as secondary variable should be preferred to minimize naturally the cross-term contribution. Therefore, for the $(\rho,\Gamma_{1})$ structural pair, some manipulations are made using the definition of the squared adiabatic sound speed, $c^{2}$, and permutations of integrals to define the target functions of the inversion. In this form, the acoustic radius inversion has a non-zero target function for the cross-term contribution. 

The target functions are defined as
\begin{align}
\mathcal{T}_{\tau,\mathrm{avg}}=& \frac{1}{2 c \tau}-\frac{m(r)}{r^{2}}\rho \left[\int_{0}^{r}\frac{1}{2 c \tau P}dx \right] \nonumber \\ & - 4 \pi r^{2} \rho \left[ \int_{r}^{R}(\frac{\rho}{x^{2}}\int_{0}^{x}\frac{1}{2 c \tau P}dy) \right]dx, \label{eqtargetavgtau} \\
\mathcal{T}_{\tau,\mathrm{cross}}&=\frac{-1}{2 c \tau}, \label{eqtargetcrosstau}
\end{align}
with $\rho$ the local density, $m$ the mass contained within the layer of radial position $r$, $P$ the pressure, $c$ the adiabatic sound speed, and $\tau$ the acoustic radius. An illustration of the target functions for 16CygA is illustrated the upper panels of Fig. \ref{FigTarIndic}. Moreover, the supplementary regularization constraints on the inversion coefficients shows that their sum must be equal to $-1$.

Acoustic radius inversions show a similar behaviour to mean density inversions, as the target functions are easily fitted by the kernels. They can thus be applied to a wide range of targets, but suffer from two main drawbacks. First, just as the mean density inversion, the SOLA method overestimates the precision of the inversion, meaning that the uncertainties derived from the propagation of the observational uncertainties cannot be trusted. Second, as a result of the behaviour of the target function, which results from the definition of the acoustic radius, the inversion is more sensitive to surface effects and thus difficult to apply to observed targets. This high sensitivity prohibits the use of acoustic radius inversions in practice and they have only been applied to the 16Cyg binary system in \cite{Buldgen2016}. While \cite{Buldgen2015tau} concluded that the method was robust with respect to surface changes in the models and non-adiabatic effects in the oscillation computations, further investigations are required to determine in more details the robustness of these inversions using more sophisticated tests and comparisons. 

\subsubsection{Core condition indicators}

Given the importance of constraining the core conditions of distant stars to determine reliable estimates of their ages, much effort has been devoted to define appropriate core condition indicators. Three of them were defined in recent years, with different observed targets in mind. 

The first indicator was defined in \cite{Buldgen2015tau} and based on the asymptotic expression of the small frequency separation. The idea was to determine the quantity

\begin{align}
t = & \int_{0}^{R}\frac{1}{r}\frac{dc}{dr}dr,
\end{align}

using the $(c^{2}, \rho)$ structural pair. As mentioned above, the use of this structural pair has been shown to lead to higher amplitude of cross-term contributions. The target function for this indicator is defined by
\begin{align}
\mathcal{T}_{t}=\frac{\frac{1}{r}\frac{dc}{dr}}{\int_{0}^{R}\frac{1}{r}\frac{dc}{dr} dr}.
\end{align}
Due to the difficulty of fitting the target function of this indicator, \cite{Buldgen2015tau} opted for a modified version of the SOLA method, trying to fit the antiderivative of the target function rather than the target function itself. This leads to a slight modification of the SOLA cost function with the first term being written
\begin{align}
\int_{0}^{R}\left[\int_{0}^{r}\mathcal{T}_{t}(x)dx - \int_{0}^{r}K_{\rm{avg}}(x) dx \right]^{2}dr.
\end{align}

This approach has been shown by \cite{Reese2012} and \cite{Buldgen2015tau} to show reasonably accurate results for both mean density and core condition indicator, at the expense of a lower stability since oscillatory behaviours are allowed around the target function of the inversion. Regarding the additional regularization term, the value of $k$ for the sum of the inversion coefficients is shown to be $1$. 

\begin{figure}
\centering
  	\includegraphics[trim=2 2 2 2, clip, width=0.95\linewidth]{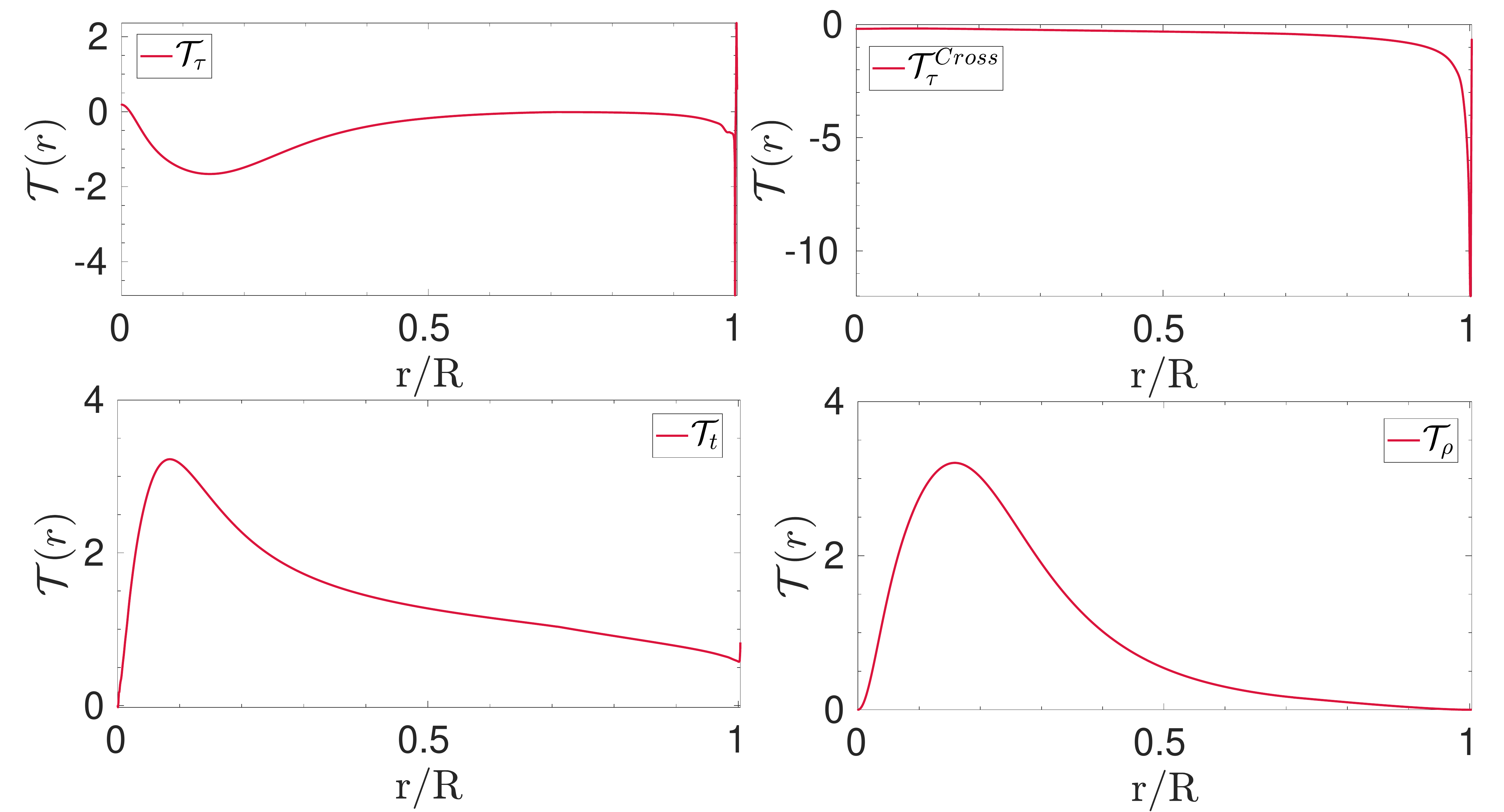}
  	   \caption{Target functions of various indicators as a function of normalized radius, taken for a model of $16$CygA. \textit{Upper left and right}: target functions for the acoustic radius inversion. \textit{Bottom left}: target function for the $t$ core condition indicator. \textit{Bottom right}: target function for the mean density.}
		\label{FigTarIndic}
\end{figure}

An illustration of the target function of the t indicator is provided in the lower left panel of Fig \ref{FigTarIndic}. \cite{Buldgen2015tau} concluded that this indicator would be mostly well-suited for young low mass stars, and published results only include tests on artificial data. It is worth noting that all indicators defined so far, namely $\bar{\rho}$, $\tau$ and $t$ have their roots in the asymptotic relations of solar-like oscillations.

\begin{figure}
\centering
  	\includegraphics[trim=2 2 2 2, clip, width=0.95\linewidth]{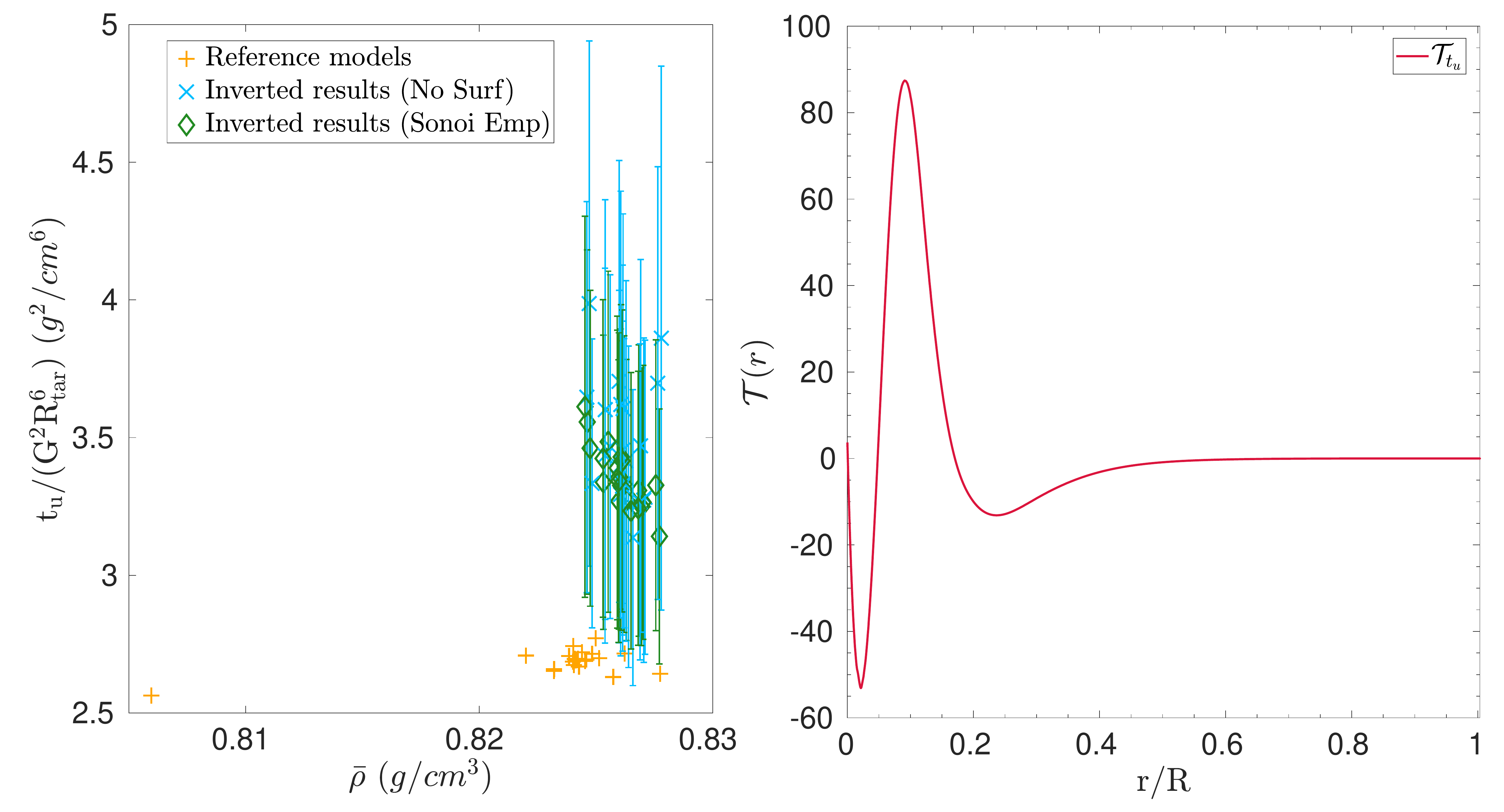}
  	\caption{\textit{Left}: $t_{u}$ inversion results as a function of mean density for 16CygA, using no surface correction (blue) and the \cite{Sonoi2015} correction (green), orange crosses showing the reference values (adapted from \cite{Buldgen2022}). \textit{Right}: target function for the $t_{u}$ indicator for a model of $16$CygA.}
		\label{Figtu16Cyg}
\end{figure}

A second core condition indicator, denoted $t_{u}$, was derived in \cite{Buldgen2015}, to allow us to access to the core conditions of low-mass stars at later stages of core hydrogen burning. The definition of this new quantity is 
\begin{align}
t_{u}=\int_{0}^{R}f(r)\left(\frac{du}{dr}\right)^{2}dr,
\end{align}
with $u=P/\rho$, the squared isothermal sound speed and $f(r)$ a weight function defined as
\begin{align}
f(r)=r (r-R) \exp \left(-7\frac{r^{2}}{R} \right), 
\end{align}
The target function for this indicator is defined as
\begin{align}
\mathcal{T}_{t_{u}}=\frac{-2 u}{t_{u}} \frac{d}{dr}\left( f(r)\frac{du}{dr} \right),
\end{align}
and the inversion is carried out using either the $(u,Y)$ or the $(u,\Gamma_{1})$ structural pair. Tests on artificial data showed that both pairs led to a similar accuracy of the inversion. From a numerical point of view, the $(u,Y)$ pair leads to a lower cross-term that is more easily damped, whereas using the $(u,\Gamma_{1})$ pair relies on the low amplitude of the relative differences in $\Gamma_{1}$ between the reference model and the target. An illustration of the target function of the $t_{u}$ indicator is shown for 16CygA in the right panel of Fig \ref{Figtu16Cyg}. The value of $k$ in Eq. \ref{eq:RegTermIndic} is of $4$ for $t_{u}$ inversions. 

The $t_{u}$ inversion has been applied to a few targets in \cite{Buldgen2016, Buldgen2016b,Buldgen2017Legacy}. In the case of 16Cyg, a recent re-study by \cite{Buldgen2022} concluded that the origin of the slight discrepancies were not due to mismatches of the internal structure. As shown in the left panel of Fig. \ref{Figtu16Cyg} illustrating the results for 16CygB, a main drawback of the $t_{u}$ inversion is its very high amplification of the observational errors, leading to a reduced significance of the inversion.

The last core condition indicator was defined in \cite{Buldgen2018} and is aimed at applications for stars with convective cores such as F-type solar-like oscillators. The idea is to carry out an inversion related to a proxy of the entropy of the stellar plasma, defined as $S_{5/3}=\frac{P}{\rho^{5/3}}$, which shows a plateau in convective regions. The height of this plateau will be sensitive to the size of the convective core. The definition of the indicator is
\begin{align}
S_{\rm{core}}=\int_{0}^{R}\frac{g(r)}{S_{5/3}}dr,
\end{align}
with 
\begin{align}
g(r)&=r\left(\alpha_{1}\exp\left(-\alpha_{2}(\frac{r}{R}-\alpha_{3})^{2} \right) + \alpha_{4} \exp\left(-\alpha_{5} (\frac{r}{R}-\alpha_{6}) \right)\right)\nonumber \\
 & \;\;\; \tanh\left(\alpha_{7}(1-\frac{r}{R})^{4}\right),
\end{align}
with $\alpha_{1}=16$, $\alpha_{2}=26$, $\alpha_{3}=0.06$, $\alpha_{4}=\alpha_{5}=6.0$, $\alpha_{6}=0.07$, and $\alpha_{7}=50$. The parameter values might be varied depending on the observed star and as discussed in \cite{Buldgen2018}. The intricate formulation of the weight is an attempt at extracting at best the information of the entropy plateau of the convective core, while keeping acceptable fits with a limited number of kernels of low degree modes. The target function of the inversion is defined in this case by
\begin{align}
\mathcal{T}_{S_{\rm{core}}}=\frac{-g(r)}{S_{\rm{core}}S_{5/3}}.
\end{align}
The value of $k$ in equation \ref{eq:RegTermIndic} is $-2/3$. The inversion will be carried out using the $(S_{5/3}, \Gamma_{1})$ or $(S_{5/3},Y)$ structural pair. 

An illustration of the target function for an F-type star model with a convective core is provided in the left panel of Fig \ref{FigTarS}, the extent of the core can be clearly seen from the plateau in the profile of $1/\tilde{S}_{5/3}$, with $\tilde{S}_{5/3}=\frac{GM^{1/3}}{R}S_{5/3}$. This inversion has been applied to artificial data in \cite{Buldgen2018}, and to actual observed targets in \cite{Buldgen2017Legacy}, \cite{Salmon2021}, and \cite{Buldgen2022}, showing in some cases significant differences with respect to the reference models. 

Detailed studies of F-type stars have not yet been undertaken and require more care as a result of the potential higher impact of surface effects. A generalization of the technique to the use of frequency separation ratios instead of individual frequencies could perhaps alleviate the issue.

\subsubsection{Envelope indicator}

In addition to defining core conditions indicators, \cite{Buldgen2018} decided to attempt at providing constraints on the entropy plateau in the convective envelope of solar-like stars. This appproach was partly motivated by results in the solar case \citep{Buldgen2017S}, showing a significant variation of the height of the entropy plateau in the solar convective zone for models built with different opacity tables. 

\begin{figure}
\centering
  	\includegraphics[trim=2 2 2 2, clip, width=0.95\linewidth]{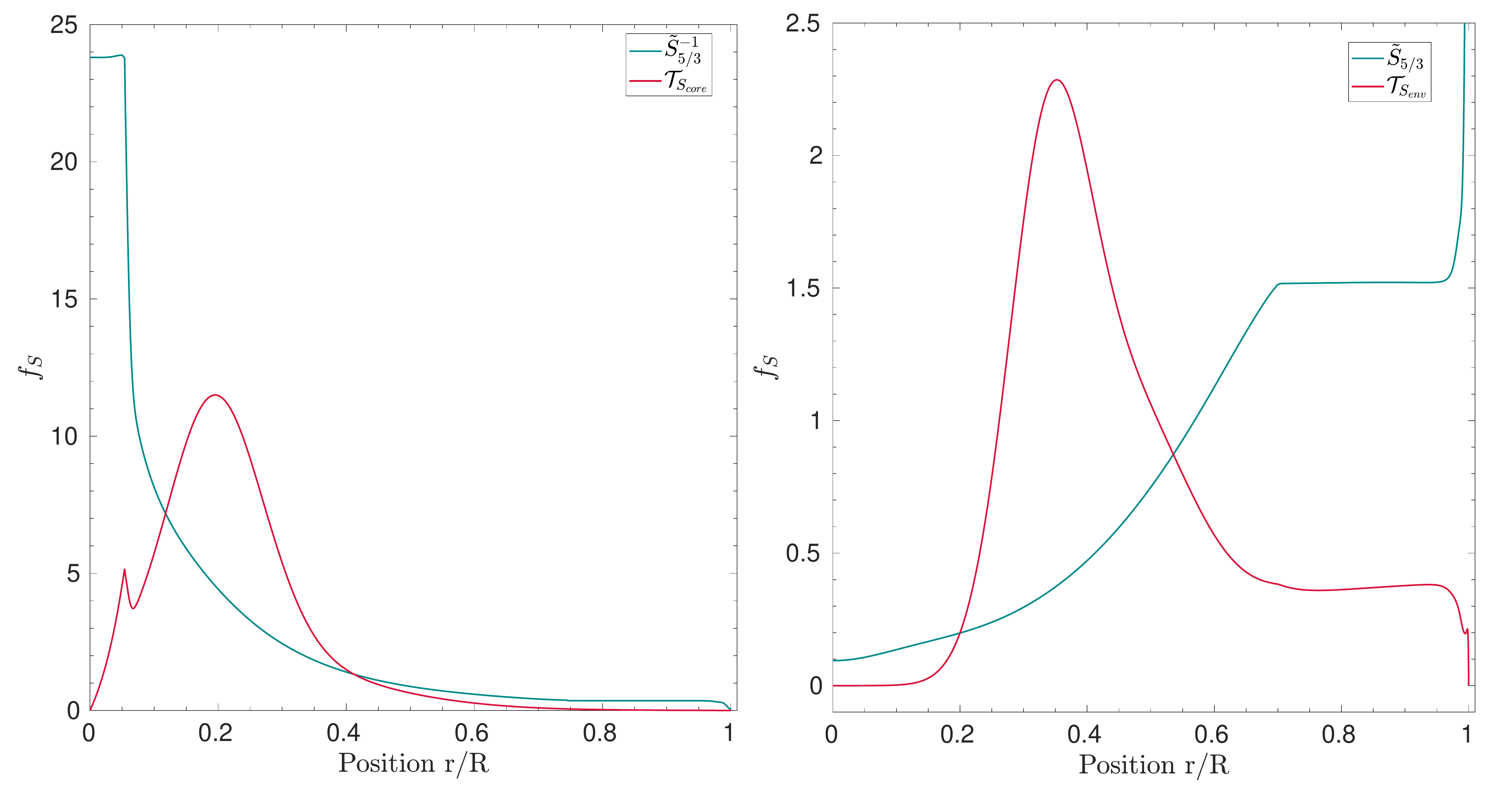}
  	\caption{\textit{Left}: target function of the $S_{\rm{core}}$ indicator and normalized inverse profile of $S_{5/3}$ as a function of normalized radius for a $1.07$M$_{\odot}$ harboring a convective core. \textit{Right}: target function of the $S_{\rm{env}}$ indicator and normalized profile of $S_{5/3}$ as a function of normalized radius for a solar model (adapted from \cite{Buldgen2018}).}
		\label{FigTarS}
\end{figure}

The indicator was defined as follows
\begin{align}
S_{\rm{env}}=\int_{0}^{R}h(r) S_{5/3}dr. \label{eq:DefSEnv}
\end{align}
with the weight function $h(r)$ defined as
\begin{align}
h(r)& =\left[\alpha_{1} \exp\left(-\alpha_{2}(\frac{r}{R}-\alpha_{3})^{2}\right)+\alpha_{4} \exp\left(-\alpha_{5}(\frac{r}{R}-\alpha_{6})^{2}\right)+\frac{0.78}{1+\left(\exp\left((\frac{R}{r}-\frac{1}{\alpha_{7}})/\alpha_{8}\right)\right)}\right]\nonumber \\ & \;\;\; \times r^{\alpha_{9}} 
\tanh \left(\alpha_{10}\left(1-\left(\frac{r}{R}\right)^{4}\right)\right),
\end{align}
with $\alpha_{1}=30$, $\alpha_{2}=120$, $\alpha_{3}=0.31$, $\alpha_{4}=7.3$, $\alpha_{5}=26$, $\alpha_{6}=0.33$, $\alpha_{7}=1.7$ $\alpha_{8}=1.2$, $\alpha_{9}=1.5$, and $\alpha_{10}=50$. 
The target function of the inversion is given by
\begin{align}
\mathcal{T_{\rm{env}}}=\frac{h(r)S_{5/3}}{S_{\rm{env}}}.
\end{align}
With these definitions, the value of $k$ in the additional regularization term is given by $2/3$. An illustration of the target function, as well as the normalized variable $\tilde{S}_{5/3}=\frac{GM^{1/3}}{R}S_{5/3}$ is provide in right panel of Fig \ref{FigTarS}.

While the number of parameters in the definition of the weight function is high, only the $\alpha_{1}$, $\alpha_{2}$ and $\alpha_{3}$ values are modified in practice. They drive the peak in the target function that can be used to extract information in intermediate radiative regions or even regions close to the base of the convective envelope. It is however extremely difficult to extract information on the entropy plateau of the convective envelope for asteroseismic targets due to the absence of high and intermediate degree modes. Thus, the corrections derived by the $S_{\rm{env}}$ indicator will be linked to a mixture of variations in the deep and intermediate radiative layers. 

This indicator has been tested on artificial data, showing promising results, and more recently on 16Cyg, where it did not succeed in providing significant corrections to the models, but rather a confirmation of their quality. This indicator seems to be better suited for low mass stars, for which the convective envelope goes deeper and can be more easily probed with low degree modes using the variational expressions. 

\section{Surface independent non-linear inversions}\label{Sec:NonLinear}

As noted above, linear variational inversions show two main weaknesses. First, the simplified single step correction provided by linear methods may not be sufficient to actually reproduce the data. Iterative approaches (although used for solar models \citep{Antia94Nonlin}, but have yet to be successfully adapted and applied to asteroseismic data. The adaptation would be quite difficult, especially with the RLS method originally used that usually retains a strong linear trend due to the regularization term. In contrast, damping too much the regularization leads to non-physical oscillatory behaviour of the solution, which is also problematic. Switching from the RLS method to the SOLA method, as done for example in \citet{Buldgen2020} for the Sun, would require additional interpolations and the convergence of the technique would depend heavily on the behaviour of the averaging kernels as well as on the surface correction. 

 Indeed, the use of individual frequencies leads to a strong impact of  surface effects on the inferred values, meaning that the inversion will be intrisically limited by the accuracy of the empirical surface corrections. Using relative differences in frequency separation ratios instead of individual frequencies may partially alleviate the issue, but it might lead to intrinsic non-linearity problems and frequency separation ratios might still be, in some cases, significantly affected by activity effects \citep{Thomas2021}. Another limitation of using frequency separation ratios is that the information of the stellar mean density is lost through the scaling of the ratios by the large frequency separation. 

In this context, asteroseismic inversions would ideally need to be able to intrisically and reliably circumvent the surface effects of solar-like oscillations, within an efficient iterative scheme. 

In the following Sections, we will present two approaches based on the phase relation of solar-like oscillations that fulfill those two requirements. Another advantage of these methods is that, as they solve the full fourth order system of oscillation equations, the issue of mode coupling arising in the case of mixed modes observed in subgiants and red giants will not affect the results, unlike the variational equations \citep{Ong2020, Ong2021}.

The approach chosen here is based on the fact that if $\omega$ is an eigenfrequency of the star, then it must satisfy a phase equation \citep{Vorontsov1998,Roxburgh2000} in which the internal structure can be described from inner phaseshifts and the outer layers from outer phaseshifts (see below). In such an approach, the fitting is carried out by reconnecting partial wave solutions in the inner and outer layers at a suitable point. The reconnexion point is chosen such that the oscillation will be almost vertical but still deep enough so that surface effects do not impact the inner solution. 

This method allows us to efficiently separate the contribution of the outer layers without the need for additional empirical corrections. As the inner phase shift is independent of the surface layers and is the constraint used to represent the internal structure, the inversion of the inner layers is essentially independent of the surface effects.  

The technique will be applied to HD177412A, the more massive component of the binary system HIP 93511, using the dataset derived by \cite{Appourchaux2015} from almost two years of continuous Kepler observations.

\subsection{Non-linear inversion using inner phaseshifts} \label{Sec:Phaseshifts}

The inversion technique presented in this Section subtracts the effect of the surface layers and seeks to infer the structure of the inner layers. More specifically, we use as fitting condition that the inner phase shifts of the solution of the oscillation equations of a model, using the observed frequencies, should collapse to a function only of frequency in the outer layers of a star \citep{Vorontsov1998,Vorontsov2001,Roxburgh2003Diff,Roxburgh2015}.  In other words, we use the fact that the contribution of the outer layers is a function only of frequency (although unknown) for low degree modes.
Other surface layer independent fitting procedures could equally be used, e.g. matching the ratio of small to large separations of the model and observations, or matching phase differences \citep{Roxburgh2013,Roxburgh2015}.

The inner phase shift $\delta_{n,\ell}(\nu)$ is the departure of the solution of the oscillation equations from a harmonic function and is defined by
\begin{align} 
\frac{\omega\psi}{d\psi/dt} = \tan(\omega \tau - \ell\pi/2 +\delta), ~~~\mathrm{where}~~~ \psi(t)= \frac{r P^{'}}{(\rho c)^{1/2}},~~~~\omega=2\pi\nu, \label{eq:PsiEq}
\end{align} 
calculated at some fractional radius $x_{f}$ in the outer layers, where $\tau =\int_{0}^{r} dr/c$ is the acoustic radius at the radial position $r$, $c$ the local sound speed and $P^{'} $ the Eulerian pressure perturbation.

This provides a fitting criterion for the inversion of the form 
\begin{align}\alpha_{n,\ell} \equiv \pi \frac{\nu_{n,\ell}}{\Delta_0}+\delta_{n,\ell} -\pi(n+\ell/2)=\alpha_0(\nu),
\end{align}
where $\Delta_0$ is an estimate of the large frequency separation and $\alpha_0(\nu)$ is an unknown function only of frequency and not of $\ell$, depicting the contribution of the surface layers to the eigenfrequency. The outer phase shift, $\alpha_0$, is a parametrised function of $\nu$, (here we used a sum of $N_\alpha$ Chebyshev polynomials) and the associated coefficients are determined by a best fit to $\alpha_{n\ell}$.
The fitting condition is then 
\begin{align}
\chi^2_{df} = \frac{1}{N_\nu-N_\alpha} \sum \left(\frac{\alpha_{n,\ell}-\alpha_0}{\epsilon_{n,\ell}}\right)^2, \label{eq:PhsChi}
\end{align}
where $\epsilon_{n,\ell}$ is the error in the inner phase shift, $\delta_{n,\ell}$, obtained by propagating the error in the frequencies ($\sigma_{n,\ell}$) through the calculation of the inner phaseshift. $N_\alpha$ is the number of coefficients in the polynomial representation of the outer phase shift $\alpha_0$. It is worth noting that the value of $\Delta_0$ or the effects of a misidentification of radial order values for the modes has no impact on the method, as they are absorbed in the value of $\alpha_0$.

\subsubsection{Inversion procedure}

The acoustic structure of a model is determined by the profiles of the local density $\rho(r)$, the local pressure $P(r)$ and the first adiabatic exponent $\Gamma_1(r)$. The density profile provides the mass distribution and the pressure follows from hydrostatic support and a pressure surface value, denoted here $P_s$. The first adiabatic exponent can be taken either from an initial trial model or as given a fixed value of $5/3$ since the departure from $5/3$ remains very small in the stellar interior (except in massive stars). Errors in the value of the surface pressure are unimportant as they have only a very small effect on the interior solution obtained with the inversion.

The model can be parametrised in many ways. For example one can use the value of the stellar mass and radius, and the local density $\rho(r)$ on a radial mesh or, as we used here, with the values of $d\log\rho/dr$ at a set of radial mesh points together with some interpolation and integration algorithm to reconstruct the structure.

The starting point of the inversion is some initial trial (input) model. It can be in principle a very simplified depiction of the structure such as a polytrope, but will be in practice some stellar model calculated with some stellar evolution code. The parameters used in the representation of the model through its inner and outer phaseshifts are iteratively modified to reduce the value of $\chi^2_{df}$, stopping when it reaches $1.0$, meaning that the internal structure as seen by the data is correctly reproduced with the parametric profile.

\begin{figure*}
  \centering
   \includegraphics[width=15.87cm]{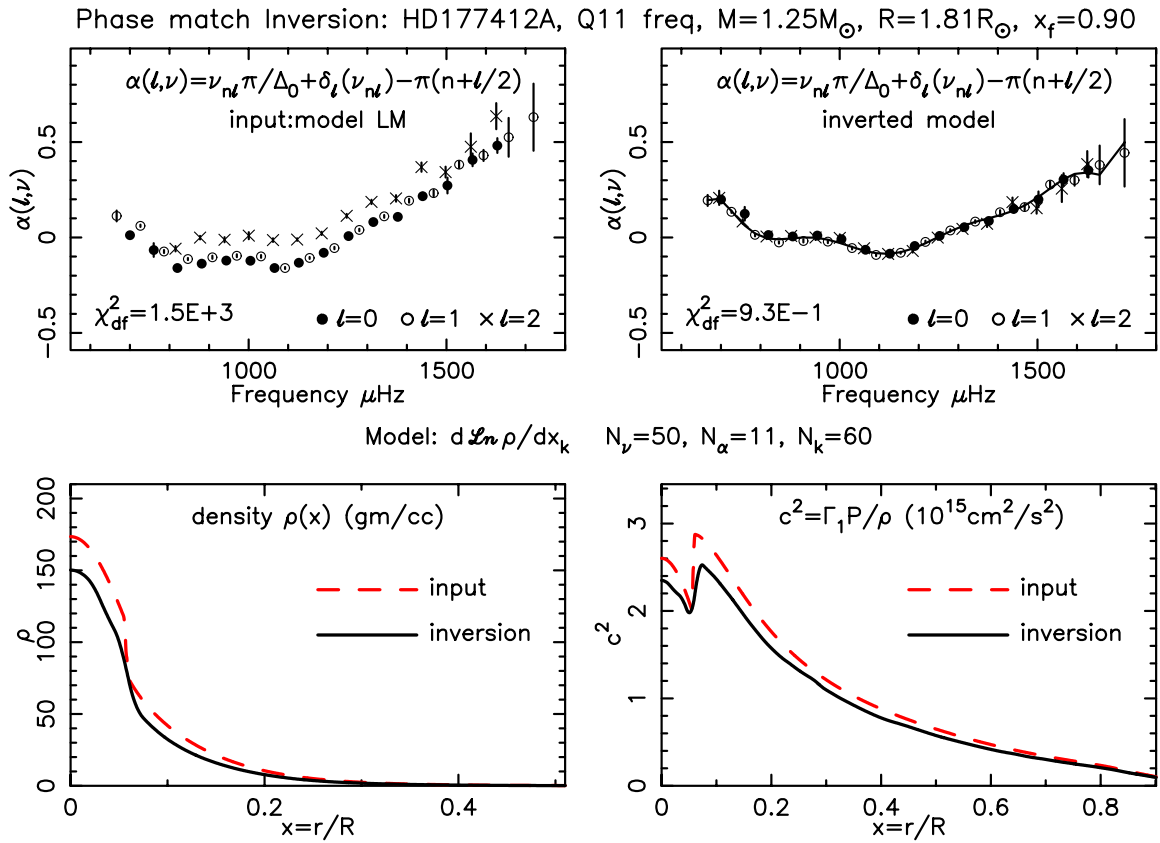}
   \caption{\textit{Upper panels}: Agreement in outer phase shift for the input ``LM'' model (left) and the inverted model (right). \textit{Lower panels}: Density (left) and squared adiabatic sound speed profiles (right) for inversion result (black) and input model (red) for the so-called ``LM'' model.}
   \label{Figirsv1}
\end{figure*}

Fig \ref{Figirsv1} shows the results starting with an input model of HD177412A computed with the Liège evolution code \citep{Scuflaire2008}, denoted here ``LM'' model. This model has a mass, M, of $1.3 M_{\odot}$, a radius, R, of $1.68 R_{\odot}$ and a central hydrogen abundance, $X_{c}$ of $0.09$. Its initial chemical composition is $ X_{0}=0.7, Z_{0}=0.02$. This model was scaled to $M=1.25 M_{\odot}$, $R=1.81 R_{\odot}$, which are the values determined from the seismic scaling relations \citep{Brown1991, Kjeldsen1995} and the observed values of $\nu_{max}$ and $\Delta$.  It should be noted that the initial model exhibited a convective core and that the inversion has kept it in the final solution.

\begin{figure*}
  \centering 
   \includegraphics[width=15.87cm]{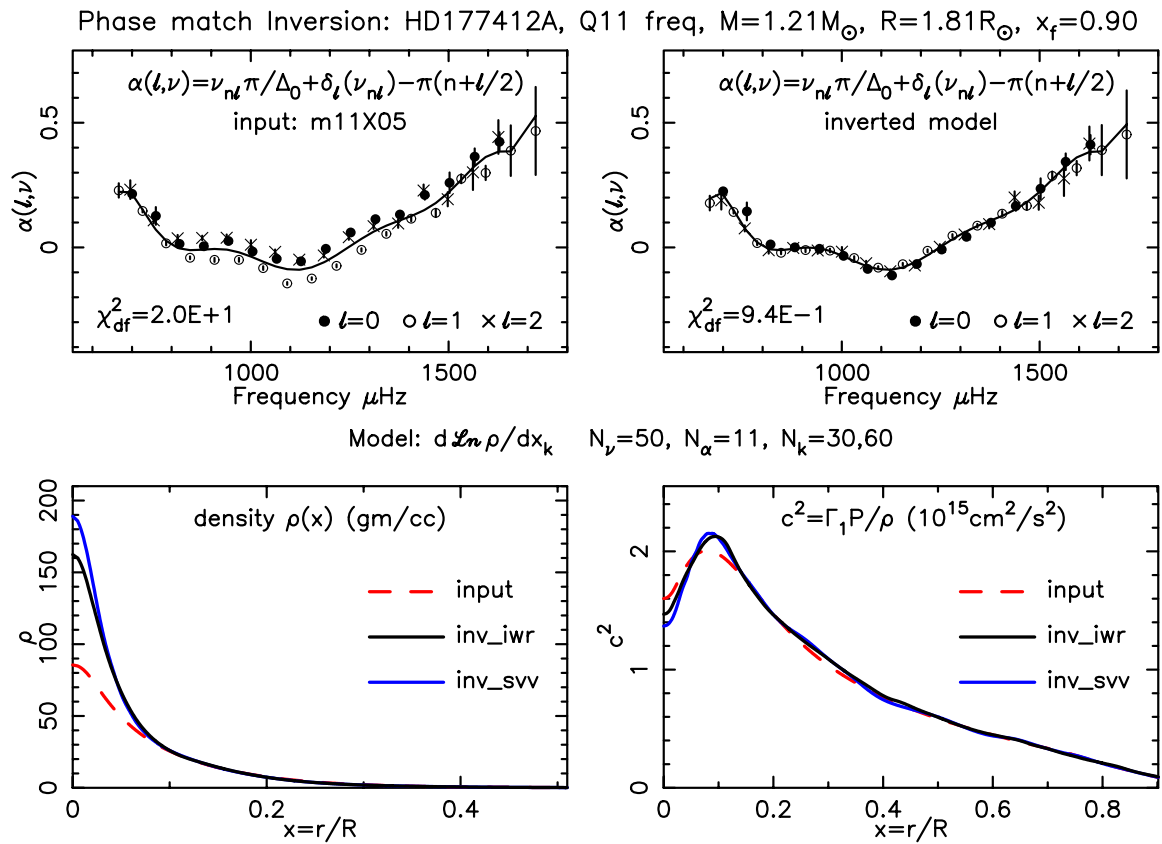} 
   \caption{Inversion result (black) and input model (red) for the so-called ``IR'' model. The profiles resulting from the iterative regularization inversion of Sect. \ref{Sec:IterativeReg} are shown in blue for comparison.}
   \label{Figirsv2}
\end{figure*}

Fig \ref{Figirsv2} shows an inversion starting with an input STAROX model \citep{Roxburgh2008} denoted here ``IR'' with a mass of $1.1M_{\odot}$, a radius of $1.22 R_{\odot}$, a central hydrogen abundance of $0.05$ and an initial chemical composition of $X_{0}=0.7, Z_{0}=0.02$. This model was also rescaled to the mass and radius values provided by the seismic scaling relations, namely $M=1.25 M_{\odot}$, $R=1.81 R_{\odot}$.

The models were represented by $N_k$ values of $d\log\rho/dr$ on a discrete mesh, and inversions undertaken with different values of $N_k$ between 10 and 180.  A downhill simplex method \citep{NeldMead65} was used to search for a minimum in $\chi^2$. The minimisation was made using the downhill simplex method with the adjustable parameters being the density derivatives on the model mesh (but any other model parameters would be suitable). The model was constructed from these model parameters - and the inner  phase shift calculated for the model using the observed frequencies as in Eq. \ref{eq:PsiEq} and chi2 as in Eq. \ref{eq:PhsChi}. Various values of the reconnexion point $x_{f}$ and different $\Gamma_{1}$ profiles (constant or that of the input model) were tested to see the impact on the results.

The results of the inversion using the iterative regularization procedure of Sect. \ref{Sec:IterativeReg} are also presented in Fig. \ref{Figirsv2}. In this case, the radius of the model was further adjusted to $1.742 R_\odot$ to improve the fit. The solution of both the method illustrated here and that of Sect. \ref{Sec:IterativeReg} are very similar, if starting from the same input model. However they differ widely from the results using the ``LM'' model instead of the ``IR'' one, as the presence of the convective core is not confirmed in the latter case.

\subsubsection{Results}
 
As is clear from the figures, different types of inverted models were obtained when starting from different initial conditions: the solution is thus not unique and it appears that some feature of the initial model remain after the inversion. This indicates a degree of degeneracy in the solutions. We can separate two families of solutions:

\begin{enumerate} 
\item Using the ``LM'' model as input, the solution of the inversion shows clear signatures of a convective core. However, as shown in Fig. \ref{Figirsv1} it was already present in the initial model. 

\item Using the ``IR'' model as input, the solution has no convective core but a steep hump in the sound speed typical of the end of main sequence evolution of lower mass stars. Again, this feature was already present in the reference model and the inversion has just enhanced it.
\end{enumerate}

We conclude that seismic inversion alone is unable to distinguish between the solution with or without a convective core. More precise data and higher frequency modes could help lift this degeneracy and distinguish between the two families of solutions. This requires a detailed study of the dependence of the inner  phase shifts on the precision of  both the model  and of the frequencies, and  possibly estimates of the limits on the luminosity of the inverted model compared with the observed value.

\subsection{Iterative regularization}\label{Sec:IterativeReg}

  The basic approach to make asteroseismic inversions insensitive to the structure of the outermost stellar layers and physics of the oscillations in these layers is here the same as in the previous section. Due to the  low values of the sound speed in the subsurface layers, the radial wavenumber of low-degree p-modes is much larger than the horizontal wavenumber. Therefore the radial eigenfunctions are expected to depend on the oscillation frequency $\omega$ only, but not on the degree $\ell$.

 The major difference is in the regularization technique of the inversion. Here, we use a nested iterative algorithm: inner and outer iterations. Inner iterations are performed with a limited number of linearized descents using conjugate gradients; the number of inner iterations plays the role of regularization parameter. The seismic model is taken as a new initial guess, and the optimization process is repeated in the outer iterations. In stellar seismology, the algorithm was tested on artificial data in \cite{Rox2002a,Roxburgh2003Diff}. It was applied to the inversion of observational frequencies of HD177412A \citep{Appourchaux2015}, with the results being described later in this section.

 The practical implementation of the technique derives largely from the  helioseismic structural inversions of \cite{Vorontsov2001, Vorontsov2002,Vorontsov2013}. The hydrostatic model is described by cubic B-splines for $m(r)/r^3$ with knots distributed uniformly in $r^2$, allowing a piecewise-analytic representation of the pressure, density and density-gradient profiles. The number of splines is chosen high enough to reproduce adequately regions of rapid spatial variation (e.g. the base of the convective envelope). In stellar inversions, the radial profile of the adiabatic exponent $\Gamma_1(r)$ is taken from the initial model and remains unchanged.

 The seismic model is truncated at some level $r=r_b$ below the photosphere, where wave propagation is close to that of a pure sound wave, but not too deep for waves to remain nearly vertical ($r_b=0.99R$ in the inversion described below). For each mode in the data set, of frequency $\omega$ and degree $\ell$, we solve the fourth-order system of the adiabatic oscillation equations by a shooting technique in the interval $0\le r\le r_b$. From the two solutions regular at the center, we form the linear combination statisfying the Laplace equation for gravity perturbations in the envelope. At the truncation boundary $r=r_b$, we match this numerical solution with the wave function $\psi_p$ proportional to the Eulerian pressure perturbation (see \cite{Vorontsov2013}), and measure the ``phase propagation time'' $T_{\ell n}$ defined as
 \begin{equation}
\omega T_{\ell n}=\pi\left(n-n_{\rm int}+\frac{1}{2}\right)
  +\arctan \frac{d\psi_p/d\tau}{\omega\psi_p}\Big|_{r=r_b}, \label{Eq:TNL}
\end{equation}
with $n$ the mode radial order, $n_{\rm int}$ the number of nodes in the Eulerian pressure perturbation below the truncation boundary, and $\tau$ the acoustic depth. $T_{\ell n}$ can be interpreted as the wave propagation time between $r=r_b$ and the upper turning point. When the model fits the observational frequencies, the $T_{\ell n}$-values of all the observed modes fit an approximation
\begin{equation}
T_{\ell n}=T_1(\omega),
\end{equation}
where $T_1(\omega)$ is a slowly-varying function of frequency. We approximate this function by a polynomial. The degree of the polynomial shall be significantly smaller than the number of modes in the observational data set, but high enough to absorb variations coming from e.g. HeII ionization (which enter $T_1(\omega)$ if this region is not well described by the model). In the results presented below, the polynomial degree is 10.

The mismatch between the model and the data is measured by the merit function $M$ defined as
 \begin{equation}
M^2={1\over N}\sum\limits_{\ell,n}
  \left[{T_{\ell n}-T_1(\omega)\over\delta T_{\ell n}}\right]^2,
\end{equation}
where $N$ is number of modes in the data set, and uncertainties $\delta T_{\ell n}$ are calculated from the uncertainties of the frequencies. Values of $M$ close to 1.0 or below indicate that the model fits the oscillation frequencies adequately.

The relations between small variations of the model parameters and variations of the phase of the wave function $\psi_p$ at the truncation boundary (the last term in Eq. \ref{Eq:TNL}), needed for the linearized descents, as well as the relation between this phase variation and variation of frequency, needed for calculating $\delta T_{\ell n}$, stem from the linear perturbation analysis of \cite{Vorontsov2013}, Appendix A.

A set of ``elementary'' model variations is then defined as partial sums of the B-splines describing $m(r)/r^3$ starting from the stellar center (i.e. each elementary variation is defined as a truncated representation of the equilibrium model). They are then normalized such as to ensure nearly equal relative variations of equilibrium density. We thus arrive to an algebraic system
\begin{equation}
A{\rm\bf x}={\rm\bf f}_\delta,
\end{equation}
where the components of the ${\rm\bf x}$ vector are the amplitudes of elementary variations, and the mismatch between $T_{\ell n}$ and $T_1(\omega)$ defines the components of ${\rm\bf f_\delta}$. The equations are then normalized to bring random errors in the right-hand side to unit variance.

The amplitudes of elementary variations are controlled by a set of orthogonal polynomials of a discrete (integer) variable (index $i$ of $x_i$). A choice of the polynomial set is known to be governed by a choice of the weight function in their orthogonality relation. This function is specified by the Euclidean norms of corresponding columns of matrix A  \citep{Strakhov2001}. This particular choice of the weight function, which defines the polynomials, ensures that the response of the components of ${\rm\bf x}$ to random frequency errors is nearly uniform, at least in the first gradient descents. The upper degree of the polynomial set has to be high enough to allow proper resolution of the inversion (set at 35 in the results below; the exact choice is not important as regularization is performed by limiting the number of iterative descents, not by constraining the functional space of allowed solutions). To ensure better stability of the inversion in the outer layers, the polynomials were additionally apodized with a cosine bell function in the interval between $r=0.7R$ and $r=0.9R$ and set to zero above.

An important ingredient of the seismic inversion when the stellar mass $M$ and radius $R$ are not well known is the degeneracy of the oscillation frequencies with respect to an homology rescaling.
 When represented in dimensionless variables, one particular seismic model describes a two-parametric family of physical models, where the density profile $\rho(r)$ scales as $M/R^3$, the squared sound speed $c^2(r)$ scales as $M/R$, the squared buoyancy frequency $N^2$ and squared oscillation frequencies $\omega^2$ both scale as $M/R^3$. Thus an initial proxy model represented in dimensionless variables will describe a two-parametric family of physical models which differ in $M$ and $R$. We bring the measured frequencies to their dimensionless values without imposing the stellar mass and radius: instead, we adjust $M/R^3$ in this scaling such as to achieve the best performance of the inversion (the best likelihood of the result after convergence). In this way, the inversion provides a best-fit value for $M/R^3$. The inverted dimentionless model now describes a one-parametric family of physical models, all of which satisfy the input data. Each model in this family can be rescaled to different values of $M$ and $R$,  keeping $M/R^3$ unchanged. In this re-scaling, which does not change the oscillation frequencies, $\rho(r)$ and $N^2(r)$ remain unchanged, but $c^2(r)$ re-scales in proportion to $M^{2/3}$.

\begin{figure}
\centering
\includegraphics[width=0.55\linewidth]{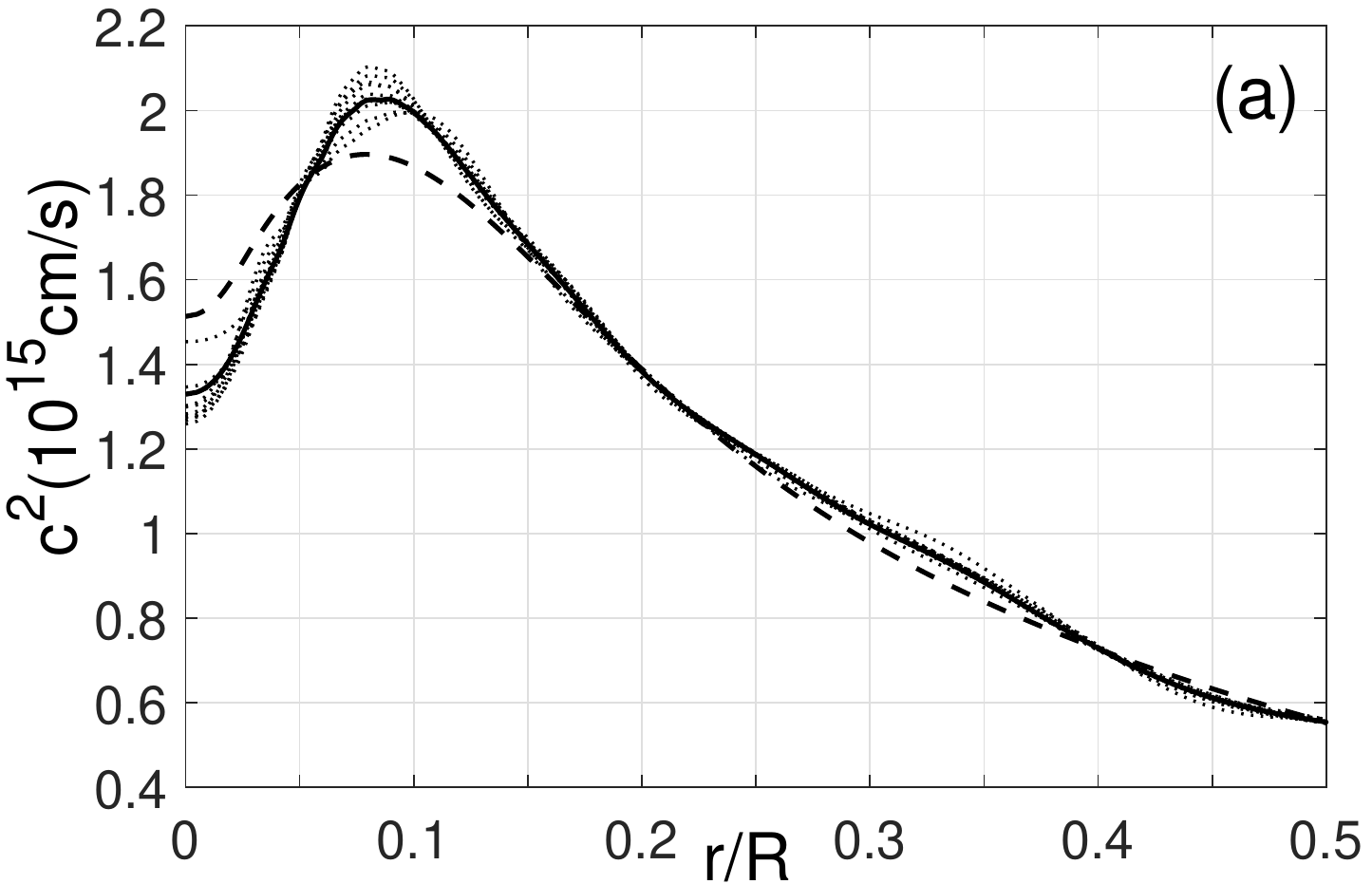}
\includegraphics[width=0.55\linewidth]{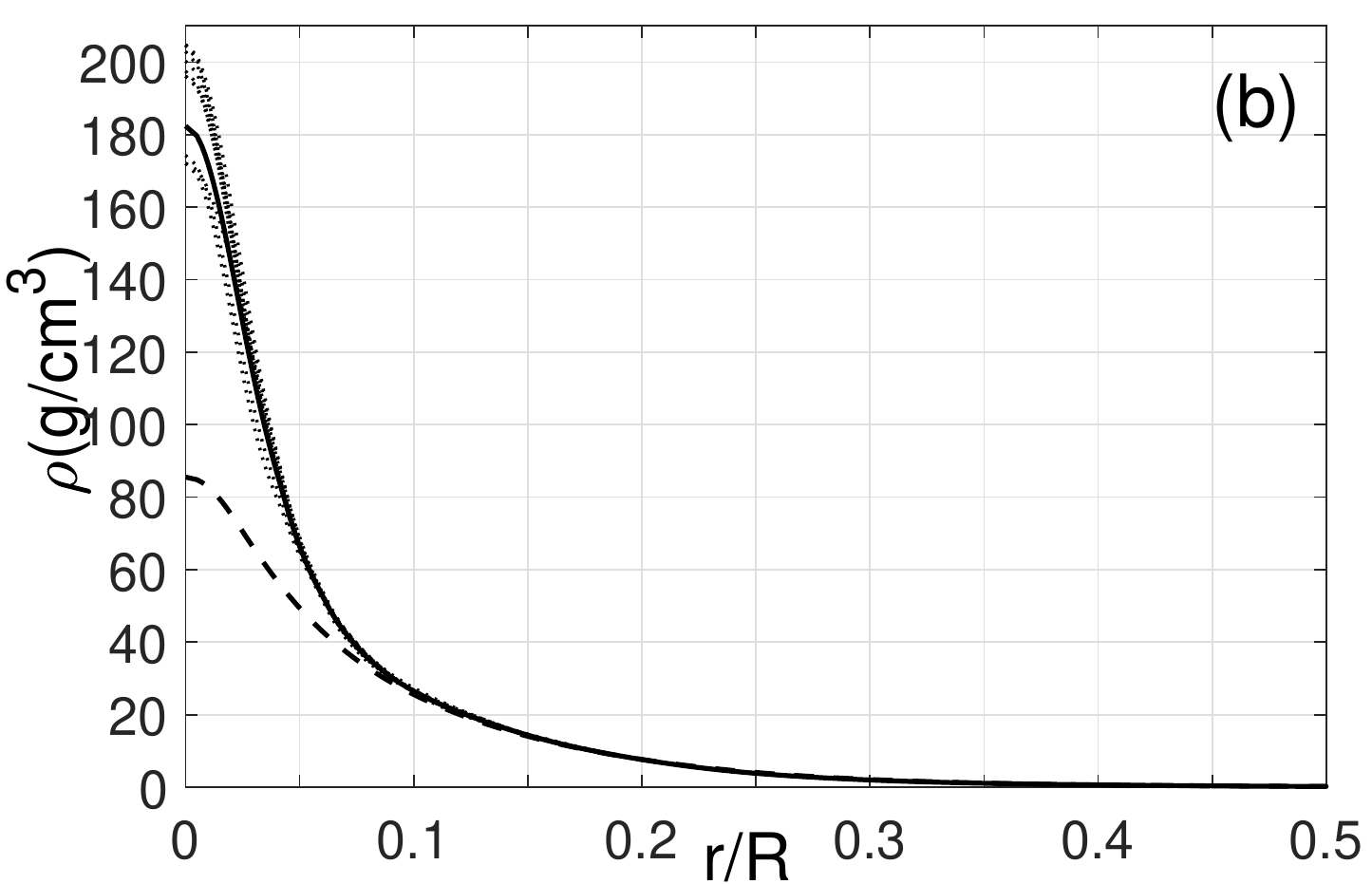}
\includegraphics[width=0.55\linewidth]{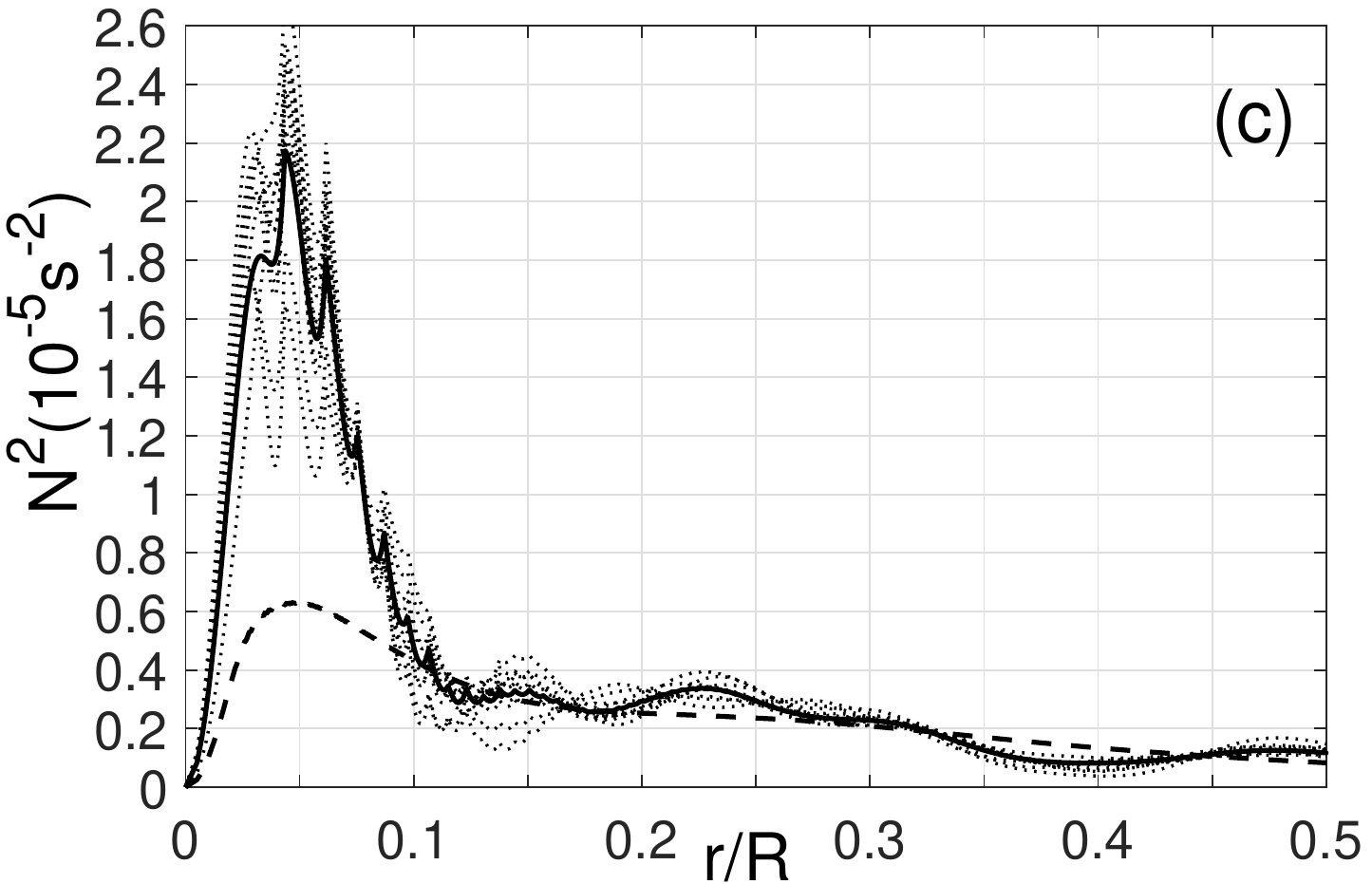}
\caption{Result of structural inversion for HD177412A: the sound speed (a), density (b) and buoyancy frequency (c). Dashed lines show a (re-scaled) model taken as an initial guess; solid lines display the inverted result. Dotted lines are the results obtained when the measured frequencies were added with Gaussian noise, of variance corresponding to the reported uncertainties, in 10 realizations, to address the sensitivity of the inversion to random errors in the input data. The nearly-optimal mean density of the model is $0.293g/cm^3$. The $c^2$-scale corresponds to a model of the original $M=1.1M_\odot$ mass but a bigger $R=1.74R_\odot$ radius.}
\label{fig:IterInv}
\end{figure}

 The results obtained with the observational p-mode frequencies of HD177412A and an evolved model of $1.1M_\odot$ star with a central hydrogen abundance of $X_c=0.05$ (model ``IR'' of the previous section) taken as an initial guess are shown in Fig. \ref{fig:IterInv}. Descents to an adequate value of the merit function (below 1.0) were performed in 7 inner and 15 outer iterations. The steep decrease in the sound-speed towards the center (panel a of Fig. \ref{fig:IterInv}), together with big density contrast (panel b of Fig. \ref{fig:IterInv}) indicate that the star is at a very late stage of the main-sequence evolution. The resulting steep gradient in the molecular weight is responsible for the sharp variation of the buoyancy frequency (panel c of Fig. \ref{fig:IterInv}). We note that the prominent wiggles in the $N^2$-curves below $0.1R$ are due to model discretisation (the cubic spline for $m(r)/r^3$ is continuous together with two derivatives, but provides $N^2(r)$ with discontinuities in its gradient).

We do not see any signature of a convective core in the results of this inversion.
Comparing with results reported in the previous section, we have to admit that with the amount and quality of frequency measurements available for HD177412A, the seismic inversion alone cannot answer the question of whether or not the star has a convective core. Additional non-seismic constraints have to be invoked to address this question.

\section{Conclusion}\label{Sec:Conclusion}

This paper is an attempt at providing a brief review of the available inversion techniques for determining the structure of solar-like oscillators. In general, seismic inversions can also involve the determination of the internal structure of a star from evolutionary computations or static computations, which we briefly discussed in Sect. \ref{Sec:Modelling}. Most inversions wrongly refer today to the specific class of the linear methods based on the variational integral relations between frequency perturbations and structural corrections. These techniques have originated in helioseismology and are now being applied to the high-quality datasets provided by space-based photometry missions. 

These methods still allow us to provide interesting estimates of the corrections to be applied to the internal structure of a given target, but intrinsic limitations such as the linearity of the integral relations and the treatment of surface effects remain major weaknesses of these techniques. In practice, while the era of space-based photometry missions has provided high quality data, it is still far from being enough to enable a full scan of the internal structure of a distant star using, for example, the SOLA method described in Sect. \ref{Sec:LinearMethods}. As discussed in this section, the versatility in the definition of the target function of the SOLA method allows us to compromise between determining global or local estimates of corrections, with global estimates being sometimes more easy to derive from a limited dataset. Another good example of this versatility being shown in \cite{Pijpers2021}. Such approaches have been applied with success to the best \textit{Kepler} targets and will remain applicable to both TESS and PLATO data in the future, especially to very precisely estimate the mean density of observed targets from a limited oscillation spectrum. Inversions of indicators as well as localized inversions would remain only applicable for the best targets with very rich oscillation spectra. 

Further improvements of the linear methods include the development and application of inversions based on relative frequency separation ratios. Indeed, \cite{Deheuvels2016} and \cite{Farnir2019} have shown that they could be used to constrain the extent of convective cores from solar-like oscillations and such inversions would likely alleviate the issue of the surface-effect dependency. 
 
In addition to discussing linear variational inversion techniques, we also presented surface independent non-linear inversions based on the phase shifts of solar-like oscillations. Such methods have proven to be very efficient at determining the internal structure of solar-like oscillators for both artificial and real \textit{Kepler} datasets while suppressing efficiently the contribution of surface regions. 

In practice however, our results confirm that non-seismic data can play a key role in discriminating between various families of inverted models. In this context, GAIA data will certainly be helpful in determining precise luminosity values, provided that accurate spectroscopic parameters are available. Interferometric radii determinations, when available, may also prove extremely helpful in this respect. It is indeed no surprise that the results presented here are for bright components of binary system, which are known to be prime testbeds of the theory of stellar structure and evolution.

An important point to note is that most of the inversions so far have been performed for main-sequence solar-like oscillators exhibiting only pure pressure modes (with the exception of \cite{Kosovichev2020} and \cite{Bellinger2021}). The wealth of seismic information contained in mixed oscillation modes thus still remains to be fully exploited. In this aspect, the non-linear inversions presented in Sect. \ref{Sec:NonLinear} are a promising avenue to take directly into account the intrinsic non-linearity of the mixed modes and to provide constraints on the internal stratification of subgiant and red-giant stars. Initial applications of phase matching for mixed modes can be found in \cite{Roxburgh2015} for artificial data and a detailed characterization using evolutionary models of the stratification of a subgiant can be found in \cite{Noll2021}, studying its consequence for core overshooting in the main sequence. These works provide benchmark approaches to further constrain the internal structure of evolved stars, for which insights on the core stratification will play a key role in improving our understanding of the missing efficient angular momentum transport at play in these stages. 

\section*{Conflict of Interest Statement}

The authors declare that the research was conducted in the absence of any commercial or financial relationships that could be construed as a potential conflict of interest.

\section*{Author Contributions}

G.B., J.B. and D.R.R. provided the data for Sections 2, 3, 4 and 5. I.W.R. and S.V.V. provided the data for Section 6. All authors read the manuscript and contributed to the discussion.  

\section*{Funding}
G.B and J.B. acknowledge fundings from the SNF AMBIZIONE grant No 185805 (Seismic inversions and modelling of transport processes in stars).

\section*{Acknowledgments}
This is a short text to acknowledge the contributions of specific colleagues, institutions, or agencies that aided the efforts of the authors.

\section*{Supplemental Data}
 \href{http://home.frontiersin.org/about/author-guidelines#SupplementaryMaterial}{Supplementary Material} should be uploaded separately on submission, if there are Supplementary Figures, please include the caption in the same file as the figure. LaTeX Supplementary Material templates can be found in the Frontiers LaTeX folder.

\section*{Data Availability Statement}
The datasets [GENERATED/ANALYZED] for this study can be found in the [NAME OF REPOSITORY] [LINK].

\bibliographystyle{Frontiers-Harvard}
\bibliography{ReviewInv}

\end{document}